\documentclass[orivec,runningheads,letterpaper]{llncs} 

\usepackage{graphicx}
\usepackage[tight,footnotesize]{subfigure}
\usepackage{tikz}
\usetikzlibrary{arrows,chains,matrix,positioning,scopes}
\usepackage{amsmath}
\usepackage{amssymb}
\usepackage{enumerate}
\usepackage{bbm}
\usepackage{dsfont}
\usepackage{pdfpages}
\usepackage{pgfplots}
\usetikzlibrary{plotmarks}
\usepackage{multirow}
\usepackage[nodayofweek]{datetime}
\usepackage{stmaryrd}
\usepackage{wasysym}
\usepackage{makecell}

\usepackage[noend]{algorithmic}
\usepackage{algorithm}
\usepackage{url}

\newlength {\localLength}    

\newcommand{\newlengthsettowidth}[2]{\newlength {#1} \settowidth{#1}{#2}}
\newcommand{\newcounterset}      [2]{\newcounter{#1} \setcounter{#1}{#2}}

\newcommand{\ensurecommand}[2]{
  \providecommand{#1}{}
  \renewcommand{#1}{#2}}

\newcommand{\NN}{\mathds N}                             
\newcommand{\ZZ}{\mathds Z}                             

\newcommand{\range}[3][X]{\Ifthen{\Equal{#1}{X}}{\{}#2,\ldots,#3\Ifthen{\Equal{#1}{X}}{\}}} 

\newcommand{\atl}{\geq}                                 
\newcommand{\atm}{\leq}                                 
\newcommand{\st}  {\operatorname{|}}                    

\newcommand{\union}       {\mathbin        {\cup}}      
\newcommand{\intersection}{\mathbin        {\cap}}      
\renewcommand {\subset}     {\subseteq}                 
\ensurecommand{\subsetneq}  {\subsetneqq}               

\newcommand{\func}[3]{#1 \colon #2 \rightarrow #3}      

\newcommand{\angles}  [1]{     \langle #1       \rangle} 

\newcommand{\limplies}{\Rightarrow}               

\newcommand{\true}    {\mathit{true}}

\newcommand{\Ifthenelse}[3]{\ifthenelse{#1}{#2}{#3}}   
\newcommand{\Ifthen}    [2]{\Ifthenelse{#1}{#2}{}}
\newcommand{\Unless}    [2]{\Ifthen{\not {#1}}{#2}}
\newcommand{\Equal}     [2]{\equal{#1}{#2}}            
\newcommand{\Empty}     [1]{\Equal{#1}{}}

\newcommand{\False}        {\Equal{1}{2}}

\newdateformat{abbrevdate}{\shortmonthname\ \twodigit{\THEDAY}, \THEYEAR}

\providecommand{\Draftmode}{\False}

\newcommand{\Itedraft}    [2]{\Ifthenelse{\Draftmode}{#1}{#2}}
\newcommand{\Ifdraft}     [1]{\Itedraft{#1}{}}

\newcommand{\drafttext}   [2][\draftcolor]{\Ifdraft{{\color{#1}#2}}}
\newcommand{\draftpointer}{$^*$}
\newcommand{\draftmargin} [2][\draftcolor]{\drafttext[#1]{\draftpointer\marginpar[\raggedleft\small{\color{#1}#2}]{\raggedright\small{\color{#1}#2}}}} 

\newcommand{\draftauthor} [3]{
  \newcommand{#1}[1]{\drafttext  [#3]{##1}}   
  \newcommand{#2}[1]{\draftmargin[#3]{##1}}}  

\newcommand{\useifspacecommand}[1]{\draftauthor{\ifspace}{\ifspacemargin}{#1}}

\Ifdraft{\setlength{\overfullrule}{10mm}}

\newcommand{\C}{\texttt C}

\newcommand{\centerTwoOut}  [2]               {#1 \hfill #2}                          

\newlength{\posLength}
\newcommand{\pos}[3][c]{\settowidth{\posLength}{#3}\makebox[\posLength][#1]{#2}}
\newlengthsettowidth{\tabLength}{\ \ \ }

\newcommand{\wbox}[2][\tabLength]{\hspace*{#1}\mbox{#2}\hspace*{#1}}

\newcommand{\Paragraph}[1][\baselineskip]{\vspace{#1}}
\newcommand{\0}[1]{}                                    
\newcommand{\End}{\end{document}}                       
\newcommand{\format}{}                                  
\newcommand{\emphasize}[1]{\textbf{#1}}                 
\newcommand{\emphdef}[1]{\emphasize{#1}}                

\newcommand{\defFullOrAbbrev}[2]{#1} 

\newtheorem{DEF}     {\defFullOrAbbrev{Definition} {Def.}}
\newtheorem{THE}[DEF]{\defFullOrAbbrev{Theorem}    {Thm.}}
\newtheorem{LEM}[DEF]{\defFullOrAbbrev{Lemma}      {Lem.}}
\newtheorem{PRO}[DEF]{\defFullOrAbbrev{Property}   {Prop.}}

\newcounter{OLDTHE}

\newenvironment{REPEATTHE}[1]{%
  \setcounter{OLDTHE}{\theDEF}
  \setcounter{DEF}{#1}
  \begin{THE}}{%
  \end{THE}
  \setcounter{DEF}{\theOLDTHE}}

\newcounter{OLDLEM}

\newenvironment{REPEATLEM}[1]{%
  \setcounter{OLDLEM}{\theDEF}
  \setcounter{DEF}{#1}
  \begin{LEM}}{%
  \end{LEM}
  \setcounter{DEF}{\theOLDLEM}}

\newcommand{\Proof}{\textbf{Proof}}                     
\newcommand{\explain}[2][\tab]{#1 \angles{\ \mbox{#2} \ }} 
\newcommand{\eop}[1][3mm]{\eopBox \vspace{#1}}          

\newcommand{\eopBox}{~\hfill$\Box$}                     

\algsetup{indent=1.5em}

\newcommand{\plkeyword}[1]{\textbf{#1}}

\newcommand{\FOREACH}{\FORALL}

\newcommand{\code}[1]{\texttt{#1}}

 \algsetup{indent=.8em} 
\newcommand{\refCapitalOrSmall}[3]{#1#3} 

\newcommand{\refFullOrAbbrev}[2]{#1} 

\newcommand{\appendixref}   [2][!]{\genericref[#1] A a {\refFullOrAbbrev{ppendix}   {pp.}} {appendix}   {#2}}
\newcommand{\algorithmref}  [2][!]{\genericref[#1] A a {\refFullOrAbbrev{lgorithm}  {lg.}} {algorithm}  {#2}}

\newcommand{\definitionref} [2][!]{\genericref[#1] D d {\refFullOrAbbrev{efinition} {ef.}} {definition} {#2}}
\newcommand{\figureref}     [2][!]{\genericref[#1] F f {\refFullOrAbbrev{igure}     {ig.}} {figure}     {#2}}

\newcommand{\lemmaref}      [2][!]{\genericref[#1] L l {\refFullOrAbbrev{emma}      {em.}} {lemma}      {#2}}
\newcommand{\lineref}       [2][!]{\genericref[#1] L l {\refFullOrAbbrev{ine}       {ine}} {line}       {#2}}

\newcommand{\propertyref}   [2][!]{\genericref[#1] P p {\refFullOrAbbrev{roperty}   {rop.}}{property}   {#2}}
\newcommand{\sectionref}    [2][!]{\genericref[#1] S s {\refFullOrAbbrev{ection}    {ect.}}{section}    {#2}}

\newcommand{\theoremref}    [2][!]{\genericref[#1] T t {\refFullOrAbbrev{heorem}    {hm.}} {theorem}    {#2}}

\newcommand{\Propertyref}   [1]{\Genericref{\refFullOrAbbrev{Property}   {Prop.}}{property}   {#1}}

\newcommand{\equationref}[2][!]{\Ifthen{\Equal{#1}!}{\refCapitalOrSmall E e {\refFullOrAbbrev{quation}{q.}}~}(\ref{equation: #2})}
\newcommand{\Equationref}[1]                                                {\refFullOrAbbrev{Equation}{Eq.}~(\ref{equation: #1})}

\newcommand{\genericref} [6][!]{\Ifthen{\Equal{#1}!}{\refCapitalOrSmall{#2}{#3}{#4}~}\ref{#5: #6}} 
\newcommand{\Genericref} [3]                                                   {{#1}~\ref{#2: #3}} 

\newcommand{\CASE}[1]{\STATE \textbf{case} #1\textbf{:} \begin{ALC@g}}
\newcommand{\ENDCASE}{\end{ALC@g}}

\newcommand{\DEFAULT}{\STATE \textbf{default:} \begin{ALC@g}}
\newcommand{\ENDDEFAULT}{\end{ALC@g}}
\newcommand{\DEFAULTLINE}[1]{\STATE \textbf{default:} }

\newcommand{\downto}{\textbf{downto}}

\renewcommand{\refFullOrAbbrev}[2]{#2} 
\renewcommand{\defFullOrAbbrev}[2]{#2} 

\renewcommand{\bf}{\bfseries}

\renewcommand{\it}{\itshape}

\draftauthor{\thomas}{\thomasmargin}{green}
\draftauthor{\peizun}{\peizunmargin}{blue}

\newcommand{\ifspacecolor}{cyan}
\useifspacecommand{\ifspacecolor}

\newcommand{\symbolcomment}[2]{\stackrel{\mbox{\tiny{(#1)}}}{#2}}

\newcommand{\expand}  [1]{{#1}^+}
\newcommand{\quotient}[1]{\overline{#1}}

\newcommand{\familyoperator}{_} 

\newcommand{\ttdsymbol}{\mathcal P} 

\newcommand{\familyobject}[2]{\Ifthenelse{\Empty{#1}}{#2}{{#2} \familyoperator {#1}}} 

\newcommand{\ttd}      [1][]{\familyobject{#1}{\ttdsymbol}} 
\newcommand{\ttdshareds} S
\newcommand{\ttdlocals}[1][]{\familyobject{#1}{L}}
\newcommand{\ttdstates}[1][]{\familyobject{#1}{V}}
\newcommand{\ttdtranss}[1][]{\familyobject{#1}{R}}
\newcommand{\localpart} W

\newcommand{\threadstate}[2]{(#1,#2)}

\newcommand{\edge}   [5][]{\threadstate{#2}{#3} \Ifthenelse{\Empty{#1}}{    \rightarrow}{\stackrel{#1}{    \rightarrow}} \threadstate{#4}{#5}} 
\newcommand{\expedge}[5][]{\threadstate{#2}{#3} \Ifthenelse{\Empty{#1}}{\dashrightarrow}{\stackrel{#1}{\dashrightarrow}} \threadstate{#4}{#5}} 

\newcommand{\ttdexpand}      {\expand   \ttd}
\newcommand{\ttdtranssexpand}{\expand   \ttdtranss}
\newcommand{\ttdquotient}    {\quotient \ttd}

\newcommand{\subscriptobject}[2]{\Ifthenelse{\Empty{#1}}{#2}{{#2}_{#1}}} 

\newcommand{\ttdshared}[1][]{\subscriptobject{#1}{s}}
\newcommand{\ttdlocal} [1][]{\subscriptobject{#1}{l}}
\newcommand{\ttdstate} [1][]{\subscriptobject{#1}{t}}

\newcommand{\ttdinitindex}{I}
\newcommand{\ttdinitshared}  {\ttdshared[\ttdinitindex]}
\newcommand{\ttdinitlocal}   {\ttdlocal [\ttdinitindex]}
\newcommand{\ttdinitstate}{\ttdstate [\ttdinitindex]}
\newcommand{\ttdinitstatepair}{\threadstate \ttdinitshared \ttdinitlocal}

\newcommand{\ttdfinalindex}{F}
\newcommand{\ttdfinalshared}  {\ttdshared[\ttdfinalindex]}
\newcommand{\ttdfinallocal}   {\ttdlocal [\ttdfinalindex]}
\newcommand{\ttdfinalstate}{\ttdstate [\ttdfinalindex]}
\newcommand{\ttdfinalstatepair}{\threadstate \ttdfinalshared \ttdfinallocal}

\newcommand{\ttdinf}{\ttd[\infty]}
\newcommand{\states}{\ttdstates[\infty]}
\newcommand{\transs}{\ttdtranss[\infty]}
\newcommand{\transsymbol} \rightarrowtail

\newcommand{\state}[2]{(#1|#2)} 

\newcommand{\covers}{\succeq}

\newcommand{\multiset}[1]{[#1]}

\newcommand{\Loop}    {\mathcal L}

\newcommand{\BWS}   {\textsc{Bws}}
\newcommand{\CovPre}{\textsc{CovPre}}

\newcommand{\ourtool}   {\textsc{Ursula}} 

\newcommand{\BFC}   {\textsc{Mcov}}

\newcommand{\KM} {\textsc{Km}}
\newcommand{\IIC}  {\textsc{IIC}}
\newcommand{\GKM}  {\textsc{Km}}
\newcommand{\Mist}  {\textsc{Mist}}
\newcommand{\Petrinizer}  {\textrm{Petrinizer}}

\newcommand{\SATABS}  {\textsc{SatAbs}}

\newcommand{\upper}{\operatorname{\uparrow}}

\newcommand{\maxplus}   [1][]{\mathbin{\varoplus}  \Unless{\Empty{#1}}{_{#1}}}
\newcommand{\maxminus}  [1][]{\mathbin{\varominus} \Unless{\Empty{#1}}{_{#1}}}

\newcommand{\iterator}{\kappa}
\newcommand{\iterate}[2][(\iterator)]{{#2}^{#1}} 

\newcommand{\Tau}{\mathcal T}

\newcommand{\invariant}[1][]{\varphi\Unless{\Empty{#1}}{_{#1}}}

\newcommand{\compos}[1][]{\mathbin{\varodot}  \Unless{\Empty{#1}}{_{#1}}}
\newcommand{\outermostmark}{\oast}
\newcommand{\pathformula}{\textsc{Path-Summary}}
\newcommand{\LAND}{\bigwedge}
\newcommand{\relation}{\bowtie}

\newcommand{\RE}     {\mathcal E}
\newcommand{\RETWO}  {\mathcal S}
\newcommand{\RETHREE}{\mathcal T}
\newcommand{\re}     {r}
\newcommand{\concat}{}
\newcommand{\choice}{\mathbin{|}}
\newcommand{\kleene}     [1]{\iterate[*]             {#1}}

\newcommand{\omost}{\emph{o-m}}
\newcommand{\imost}{\emph{i-m}}

\begin{document}

\mainmatter  


\newcommand{\theTitle}{%
  Unbounded-Thread Reachability via \\
  Symbolic Execution and Loop Acceleration (Technical Report)}


\titlerunning{A New Approach to Unbounded-Thread Reachability}

\title{%
  \theTitle%
  \thanks{This work is supported by NSF grant no.~1253331.}
}

%
%
\author{Peizun Liu
  \and Thomas Wahl
}

\authorrunning{Peizun Liu and Thomas Wahl}

\institute{Northeastern University, Boston, United States}

%
%
\toctitle{Peizun Liu's LNCS Template}
\tocauthor{Peizun Liu based on LNCS}
\maketitle


\begin{abstract}
  We present an
  approach to parameterized reachability for communicating
  finite-state threads that formulates the analysis as a
  satisfiability problem.
  In addition to the unbounded number of threads, the main challenge for
  SAT/SMT-based reachability methods is the existence of unbounded loops in
  the program executed by a thread. We show in this paper how \emph{simple}
  loops can be accelerated \emph{without approximation} into Presburger
  arithmetic constraints. The constraints are obtained via symbolic
  execution and are satisfiable exactly if the
  given program state is reachable. We summarize loops \emph{nested} inside
  other loops using
  recurrence relations derived from the inner loop's acceleration. This
  summary abstracts the loop iteration parameter and may thus
  overapproximate. An advantage of our symbolic approach
  is that the process of building the Presburger formulas may instantly
  reveal their unsatisfiability, before any arithmetic has been performed.
  We demonstrate the power of this technique for proving and refuting
  safety properties of unbounded-thread programs and other infinite-state
  transition systems.
\end{abstract}



\section{Introduction}
\label{section: Introduction}

Unbounded-thread program verification continues to attract the
attention it deserves: it targets programs designed to run on multi-user
platforms and web servers, where concurrent software threads respond to
service requests of a number of clients that can usually neither be
predicted nor meaningfully bounded from above a priori. To account for
these circumstances, such programs are designed for an unspecified and
unbounded number of parallel threads that is a system parameter.

We target in this paper unbounded-thread shared-memory programs where each
thread executes a non-recursive, finite-data procedure. This model is
popular, as it connects to multi-threaded \C\ programs via predicate
abstraction, a technique that has enjoyed progress for concurrent programs
in recent years \cite{DKKW2011}. The model is also popular since basic
program state reachability questions are decidable, although of high
complexity: the equivalent \emph{coverability problem} for Petri nets was
shown to be EXPSPACE complete \cite{CLM76}. The motivation for our work is
therefore not to solve this problem per se, but to do so with practicable
efficiency.

Building on impressive recent advances in SMT technology, the approach we
take in this paper
is to reduce the analysis to a logical decision problem. Such reductions
are common in the context of bounded model checking, where
finite-length execution paths are translated into logical constraints whose
satisfiability indicates the reachability of some (error) condition along
the path.
In our context, we intend to strengthen this principle along two lines:
\begin{enumerate}

\item[\textbf{(A)}] we are dealing with multi-threaded programs; the
  different thread interleavings give rise to too many execution paths for
  them to be enumerated; and

\item[\textbf{(B)}] we aim at finding bugs \emph{and} soundly proving
  safety; we can thus not simply bound the path length by a constant.

\end{enumerate}

In this paper we tackle challenge \textbf{(A)} by considering an
abstraction of the given program whose \emph{single-threaded} execution
overapproximates the execution of the original program by \emph{any} number
of threads. The abstraction is surprisingly simple: it allows the single
thread to change its local state in certain disciplined ways, thereby
slipping into the role of a potential parallel thread. We can now analyze
this sequential program, without regard for interleavings.

The question whether an abstract error path can be concretized
is decided via a satisfiability problem, by \emph{symbolically executing} a
known coverability algorithm~\cite{Abdulla2010} along potential
multi-threaded error paths.
Here we face challenge~\textbf{(B)}: the given
program may feature loops, an issue that is in fact exacerbated by the
additional behavior in the abstraction. To permit unbounded symbolic
execution along paths with loops, we show how \emph{simple} loops can be
accelerated, without loss of information, into a formula that specifies how
the number of threads per local state changes during one loop iteration.
These changes can be expressed in Presburger arithmetic, the decidable
theory over linear integer operations.

Complicated \emph{loop nests} are not amenable to exact acceleration.
We summarize such nests by abstracting the iteration count for inner loops
and approximating outer loops by solving a recurrence relation. This
process introduces imprecision and the potential for spurious reachability
results.
To detect this possibility, we need to recover the iteration counts for
inner loops. Our algorithm does this in a refinement cycle whose complexity
is linear in the \emph{nesting height} of the loop arrangement. The result
is a sound
symbolic coverability method that is often also able to produce paths
witnessing error state reachability. In the absence of nested loops, the
algorithm is sound and complete.

Our algorithm can be viewed as separating the branching required in
exhaustive infinite-state searches such as Abdulla's
algorithm~\cite{Abdulla2010}, and the arithmetic required to keep track of
the number of threads per local state. Our abstract structure is loop-free
and can thus be explored path by path. Each path is symbolically executed
into a Presburger formula. The question whether the target state is
reachable along this path can then often be answered very quickly. If the
path does not connect the initial and target states, it is not even
considered for symbolic execution. Contrast this with pure search
techniques, which might explore the search space into a particular
direction, only to find that all paths in this direction dead-end.

Recent work on solving coverability questions via marking equations
\cite{ELMMN14} is, to our knowledge, one of the first \emph{practical}
attempts to tackle infinite-state reachability via SMT technology. While
very efficient, this method can reliably recognize only unreachable
instances and indeed produces many spurious answers.
In this paper we show how \emph{control flow information} present in
multi-threaded programs can be exploited to obtain distinctly more precise
symbolic encodings of the reachability problem, while retaining much of its
efficiency.

\

This submission comes with an appendix containing proofs to claims made in
this paper, and other material.

\section{Thread-Transition Diagrams and Backward Search}
\label{section: Thread-Transition Diagrams and Backward Search}

We assume multi-threaded programs are given in the form of an abstract
state machine called \emph{thread transition diagram} \cite{KKW12}. Such a
diagram
reflects the replicated nature of programs we consider: programs consisting
of threads executing a given procedure defined over shared (``global'') and
(procedure-)local variables. A~thread transition diagram (TTD) is a tuple
$\ttd=(\ttdshareds,\ttdlocals,\ttdtranss)$, where
\begin{itemize}
  \item $\ttdshareds$ is a finite set of \emph{shared} states;
  \item $\ttdlocals$ is a finite set of \emph{local} states;
  \item $\ttdtranss \subset (\ttdshareds \times \ttdlocals) \times (\ttdshareds \times
    \ttdlocals)$ is a (finite) set of \emph{edges}.
\end{itemize}
An element of $\ttdstates = \ttdshareds \times \ttdlocals$ is called
\emph{thread state}. We write $\edge{s_1}{l_1}{s_2}{l_2}$ for
$((s_1,l_1),(s_2,l_2)) \in \ttdtranss$.
We assume the TTD has a \emph{unique} initial thread state, denoted
$\ttdinitstate = \ttdinitstatepair$. \appendixref{Uniqueness of the Initial
  State} explains how this can be enforced under some very light-weight
condition. An example of a TTD is shown in \figureref{ttd}.
\begin{figure}[htbp]
  \centerline{
    \subfigure[]{\includegraphics[width=1.6in]{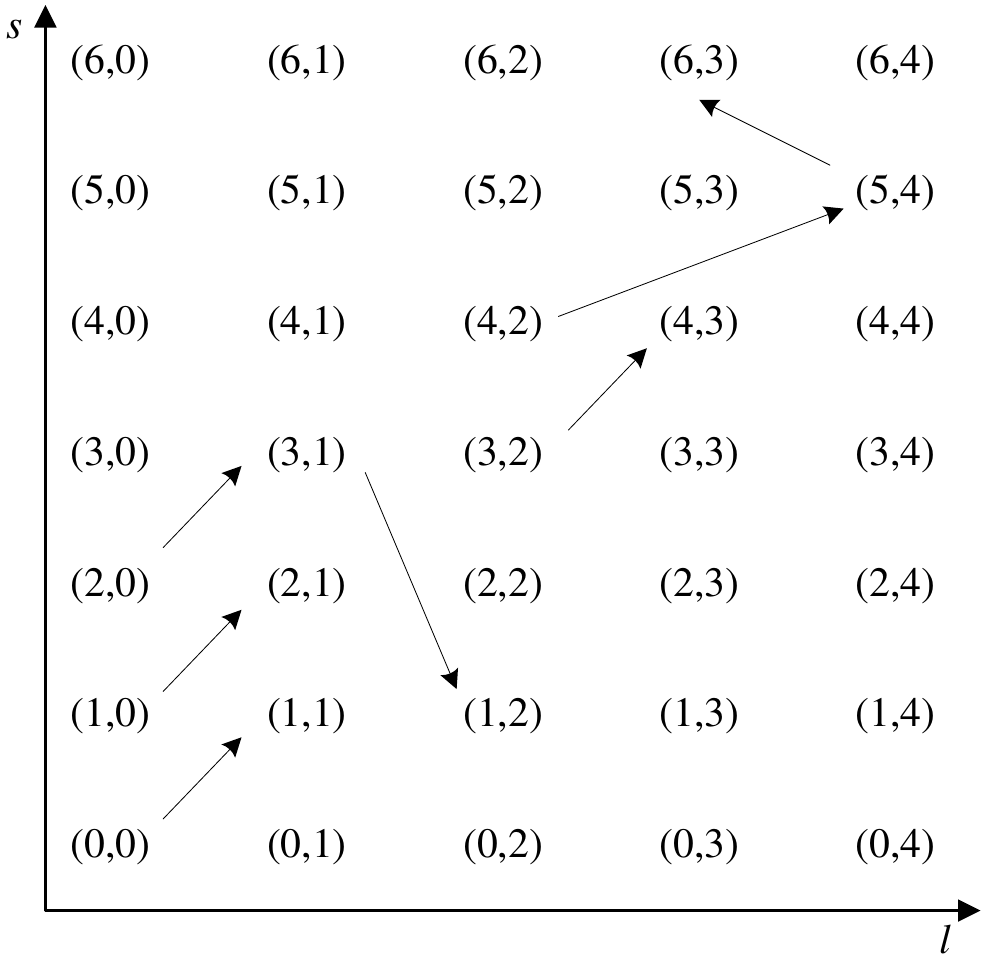}
      \label{figure: ttd}}
    \hfil
    \hspace{0.2em}
    \subfigure[]{\includegraphics[width=1.6in]{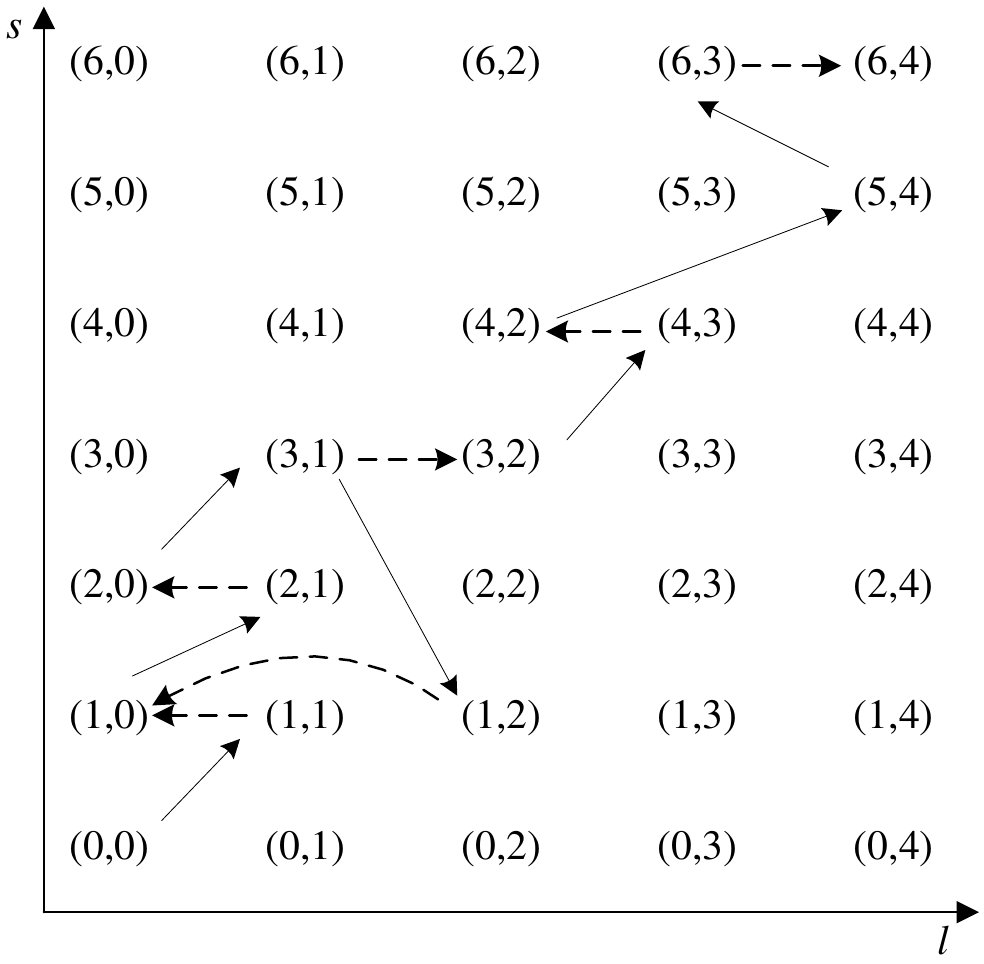}
      \label{figure: expanded ttd}}
    \hfil
    \hspace{0.2em}
    \subfigure[]{\includegraphics[width=1.35in]{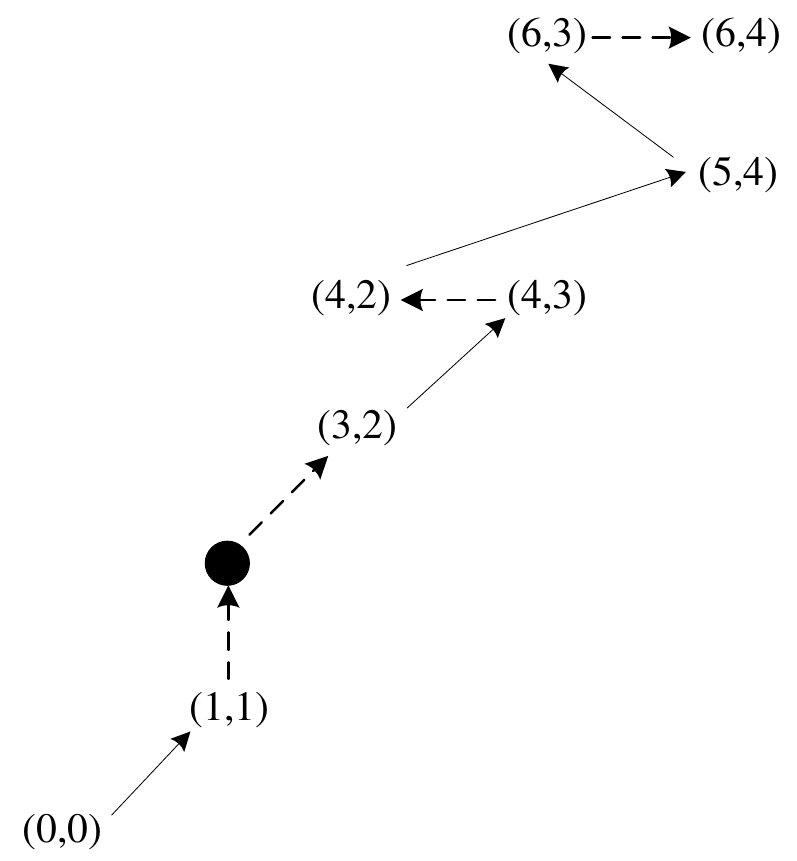}
      \label{figure: lfttd}}}
  \caption{(a) A thread transition diagram $\ttd$ (initial state
    $\ttdinitstate = \threadstate 0 0$); (b) part of the Expanded TTD
    $\ttdexpand$ with a path~$\expand \sigma$; (c) part of the SCC quotient
    graph $\ttdquotient$ of $\ttdexpand$, with quotient path $\quotient
    \sigma$. The black disc represents the loop in~$\expand \sigma$ (the
    other SCCs are trivial)}
  \label{figure: ttd abstractions}
\end{figure}

A TTD gives rise to a family, parameterized by $n$, of transition systems
$\ttd[n] = (\ttdstates[n],\ttdtranss[n])$ over the state space
$\ttdstates[n] = \ttdshareds \times \ttdlocals^n$, whose states we write in
the form $\state s {\range[]{l_1}{l_n}}$. This notation represents a global
system state with shared component $s$, and $n$ threads in local states
$l_i$, for $i \in \range 1 n$.
The transitions of $\ttd[n]$, forming the set $\ttdtranss[n]$, are written
in the form $\state s {\range[]{l_1}{l_n}} \transsymbol \state {s'}
{l'_1,\ldots,l'_n}$. This transition is defined exactly if there exists $i
\in \range 1 n$ such that $\edge s {l_i} {s'} {l'_i}$ and for all $j \not=
i$, $l_j = l'_j$. That is, our executing model is asynchronous: each
transition affects the local state of at most one thread.
The initial state set of $\ttd[n]$ is $\{\ttdinitshared\} \times
\{\ttdinitlocal\}^n$. A \emph{path} of $\ttd[n]$ is a finite sequence of
states in $\ttdstates[n]$ whose first element is initial, and whose
adjacent elements are related by $\ttdtranss[n]$. A thread state $(s,l) \in
\ttdshareds \times \ttdlocals$ is \emph{reachable in $\ttd[n]$} if there
exists a path in $\ttd[n]$ ending in a state with shared state component
$s$ and some thread in local state~$l$.

A TTD also gives rise to an infinite-state transition system $\ttdinf =
(\states,\transs)$, whose set of states/transitions/initial states/paths is
the union of the sets of states/transitions/initial states/paths of
$\ttd[n]$, for all $n \in \NN$. We are tackling in this paper the
\emph{thread state reachability question}: given a TTD $\ttd$ and a
\emph{final} thread state $\ttdfinalstatepair$, is $\ttdfinalstatepair$
reachable in $\ttdinf$~? It is easy to show that this question is
decidable, by encoding $\ttdinf$ as a \emph{well quasi-ordered system}
(WQOS)~\cite{Abdulla2010}: let the \emph{covers} relation $\covers$ over
$\states$ be defined as follows:
\[
  \state s {l_1,\ldots,l_n} \covers \state {s'} {l'_1,\ldots,l'_{n'}}
\]
whenever $s = s'$ and $\multiset{l_1,\ldots,l_n} \supseteq
\multiset{l'_1,\ldots,l'_{n'}}$, where $\multiset \cdot$ denotes a
\emph{multiset}. Relation $\covers$ is a well quasi-order on $\states$, and
\mbox{$(\ttdinf,\covers)$} satisfies the definition of a WQOS, in
particular the \emph{monotonicity} property required of $\covers$ and
$\transsymbol$. The thread state reachability question can now be cast as a
\emph{coverability problem}, which is decidable but of high complexity,
e.g.\ EXPSPACE-complete for standard Petri nets \cite{CLM76}, which are
equivalent in expressiveness to infinite-state transition systems obtained
from TTD \cite{KKW12}.


\noindent
\setlength{\localLength}{\parindent}
\begin{minipage}{70mm}
  \hspace*{\localLength}A sound and complete algorithm to decide
  coverability for WQOS
  is the \emph{backward search} algorithm by Abdulla et
  al.\ \cite{ACJT96,Abdulla2010}, a simple version of which is shown on the
  right. Input is a set of initial states $I \subseteq \states$, and a
  non-initial final state~$q$. The algorithm maintains a work set $W
  \subseteq \states$ of unprocessed states, and a set $U \subseteq \states$
  of minimal encountered states. It successively computes minimal
  \emph{cover predecessors}
  \begin{equation}
    \format\hspace*{-9pt}\CovPre(w) = \min\{p : \exists w' \covers w : p \transsymbol w'\}
    \label{equation: cover predecessors}
  \end{equation}
  starting from $q$, and terminates either by backward-reaching an initial
  state (thus proving coverability of $q$), or when no unprocessed vertex
  remains (thus proving uncoverability).
\end{minipage}
\hfill
\begin{minipage}{46mm}
  \vspace*{-7mm}
  \begin{algorithm}[H]
    \caption{$\BWS(I,q)$}
    \begin{algorithmic}[1]
      \REQUIRE{initial states $I$, \\ \quad \quad final state $q \not\in I$}
      \STMT $W := \{q\}$\,; \,$U := \{q\}$
      \WHILE{$\exists w \in W$}
        \STMT $W := W \setminus \{w\}$
        \FOR{$p \in \CovPre(w) \setminus \upper U$} \label{line: bwpreimage}
          \IF{$p \in I$}
            \STMT ``$q$ coverable''
          \ENDIF
          \STMT $W := \min(W \union \{p\})$ \label{line: discard non minimals 1}
          \STMT $U := \min(U \union \{p\})$ \label{line: discard non minimals 2}
        \ENDFOR
      \ENDWHILE
      \STMT ``$q$ not coverable'' \label{line: bc: uncoverable}
    \end{algorithmic}
    \label{algorithm: Abdulla}
  \end{algorithm}
  \vspace{-6mm}
  \footnotesize
  \algorithmref{Abdulla}: infinite-state backward search. Symbol $\upper U$
  stands for the \emph{upward closure} of $U$: \\
  $\upper U = \{u': \exists u \in U: u' \covers u\}.$
\end{minipage}

\section{Reachability as a Decision Problem: Overview}
\label{section: Reachability as a Decision Problem: Overview}

Our approach to encoding reachability in $\ttdinf$ as a decision problem
operates over an abstraction of the given TTD, with the property that any
thread state reachable in $\ttdinf$ for any number of threads is also
reachable in the abstract structure \emph{executed by a single thread}. The
search for paths to the final thread state can therefore focus on abstract
single-thread paths. The imprecision introduced by this abstraction is
eliminated later when each path is translated into a Presburger formula, as
we will see. In this section we first define this abstract structure, and
then present the intuition of our algorithm.

\subsection{A Single-Threaded Abstraction of $\ttdinf$}
\label{section: A Single-Threaded Abstraction of ttdinf}

A key operation employed during \algorithmref{Abdulla} is what we call
\emph{expansion} (of a state): the addition of a thread in a suitable local
state during the computation of cover predecessors \equationref[]{cover
  predecessors}. We can simulate the effect of such expansions
\emph{without adding threads}, by allowing a thread to repeatedly change
its local state in certain disciplined ways. To this end, we expand the TTD
data structure as follows.%
\begin{DEF}
  \label{definition: expanded ttd}
  Given a TTD $\ttd=(\ttdshareds,\ttdlocals,\ttdtranss)$, an
  \emphdef{expansion edge} is an edge $(\threadstate s l, $\format
  $\threadstate s {l'})$ (same shared state) such that $l \not= l'$. The
  \emphdef{Expanded TTD (ETTD)} of $\ttd$ is the structure
  $\ttdexpand=(\ttdshareds,\ttdlocals,\ttdtranssexpand)$ with
  $\ttdtranssexpand = \ttdtranss \cup \{e: \ \mbox{$e$ expansion edge}\}$.
\end{DEF}
To distinguish the edge types in $\ttdexpand$, we speak of \emph{real
  edges} ($\in \ttdtranss$) and expansion edges. Intuitively, expansion
edges close the gap between two real edges whose target and source,
respectively, differ only in the local state. This can be seen in
\figureref{expanded ttd}, which shows part of the ETTD generated from the
TTD in \figureref{ttd}. In the graphical representation, expansion edges
run horizontally and are shown as dashed arrows $\expedge s l s {l'}$.

As we will see, our reachability algorithm processes certain paths from
$\ttdinitstate$ to $\ttdfinalstate$, of which $\ttdexpand$ may still have
infinitely many, due to the possibility of loops. To facilitate this
process, we collapse the ETTD into a quotient structure, by replacing loops
with single nodes that represent the unique strongly connected component a
loop is part of. Let therefore $\ttdquotient$ be the (acyclic) SCC quotient
of the expanded graph $\ttdexpand$; an example is shown in
\figureref{lfttd}.

Being loop-free, the quotient graph $\ttdquotient$ contains only finitely
many paths between any two nodes. It also has another key property that
makes it attractive for our algorithm: let us interpret $\ttdexpand$ and
$\ttdquotient$ as sequential transition systems. That is, when we speak of
\emph{reachability} and \emph{paths} in $\ttdexpand$ ($\ttdquotient$), we
mean ``when $\ttdexpand$ ($\ttdquotient$) is executed by a single thread
from $\ttdinitstate$''. In contrast, in $\ttdinf$ these concepts are
interpreted over an unbounded number of threads executing $\ttd$
from~$\ttdinitstate$. Given these stipulations: $\ttdquotient$
overapproximates $\ttdinf$, in the sense that, if thread state
$\ttdfinalstate$ is reachable in $\ttdinf$,
then $\ttdfinalstate$ is also reachable in $\ttdquotient$.
This property is (proved as) part of our main correctness
\theoremref{soundness} later in this paper.

\subsection{Backward Search via Symbolic Execution}
\label{section: Backward Search via Symbolic Execution}

Given the reachability semantics defined in \sectionref{A Single-Threaded
  Abstraction of ttdinf}, each multi-thread path in $\ttdinf$ corresponds
to a single-thread path in the quotient structure $\ttdquotient$. Our
algorithm therefore first identifies paths in $\ttdquotient$ from
$\ttdinitstate$ to $\ttdfinalstate$; if none, $\ttdfinalstate$ is
unreachable in $\ttdinf$. If such paths do exist, we cannot conclude
reachability in $\ttdinf$: for example, thread state $\ttdfinalstate :=
\threadstate 6 4$ in \figureref{ttd abstractions} is easily seen to be
unreachable in $\ttdinf$, no matter how many threads execute the diagram
$\ttd$ in (a), but is obviously (sequentially) reachable in $\ttdquotient$
(c).

We therefore need to decide, for each path in $\ttdquotient$ from
$\ttdinitstate$ to $\ttdfinalstate$, whether it conversely corresponds to a
valid multi-thread path in $\ttdinf$. To this end, consider the operation
of the backward search \algorithmref{Abdulla}. Given a global state of the
form $\state{s'}{\range[]{l'_1}{l'_n}}$, it computes cover predecessors
(\equationref{cover predecessors}), by first firing edges of $\ttdtranss$
backwards whose targets equal one of the thread states
$\threadstate{s'}{l'_i}$. Second, for each edge $e$ whose target
$\threadstate{s'}{l'}$ (with shared state $s'$) does not match any of the
thread states $\threadstate{s'}{l'_i}$, \algorithmref{Abdulla} expands the
global state, by adding one thread in local state $l'$, followed by firing
$e$ backwards, using the added thread.\footnote{We exploit here the fact
  that the cover pre-image \equationref[]{cover predecessors} in systems
  induced by TTDs increases the number of threads in a state by at most 1
  (proved in \cite[Lemma~1]{LW14}).}

The steps performed by \algorithmref{Abdulla} can be expressed in terms of
updates to local-state counters. For an edge $e$ of the form $\edge s l
{s'}{l'}$, if the current global state $\state{s'}{\range[]{l'_1}{l'_n}}$
contains a thread in local state $l'$, firing $e$ backwards amounts to
decrementing the counter $n_{l'}$ for $l'$, and incrementing the counter
$n_l$ for $l$. If the current global state does not contain a thread in
local state $l'$, firing $e$ backwards amounts to temporarily incrementing
$n_{l'}$ (= setting it to~1), followed by decrementing it (= back to 0),
followed by incrementing $n_l$.

We can execute these steps \emph{symbolically}, instead of concretely, by
traversing a given path $\quotient \sigma$ in $\ttdquotient$ from
$\ttdfinalstate$ backward to $\ttdinitstate$, assuming for now we visit
only trivial SCCs. Each real edge in $\quotient \sigma$ simulates the
standard backward firing of an edge. Each expansion edge in $\quotient
\sigma$ simulates the temporary addition of a thread in a local state. We
perform these simulations by encoding the corresponding counter updates
described in the previous paragraph as logical constraints over the
local-state counters. The assertion that $\ttdfinalstate$ is reachable in
$\ttdinf$ then translates to the condition that, given these constraints,
the values for all counters at the end of the simulation, i.e.\ when
backward-reaching $\ttdinitstate$ along $\quotient \sigma$, are zero, with
the exception of $n_{\ttdinitlocal}$: this condition ensures that the
global state constructed via symbolic backward execution is of the form
$\{\ttdinitshared\} \times \{\ttdinitlocal\}^n$, i.e.\ initial.

The constraints are expressible in \emph{Presburger} (linear integer)
arithmetic. To demonstrate this, we introduce some light notation. For $x,y
\in \ZZ$ and $b \in \NN$, let $x \maxplus[b] y = \max\{x + y,b\}$.
Intuitively, $x \maxplus[b] y$ is ``$x + y$ but at least $b$''. When $b=0$,
we omit the subscript. We also use $x \maxminus[b] y$ as a shorthand for $x
\maxplus[b] (-y)$ ($ = \max\{x-y,b\}$). For example, $x \maxminus 1$ equals
$x-1$ if $x \atl 1$, and 0 otherwise.
Neither $\maxplus[b]$ nor $\maxminus[b]$ are associative: $(1 \maxplus 2)
\maxplus -3 = 0 \not= 1 = 1 \maxplus (2 \maxplus -3)$. We therefore
stipulate: these operators (i) associate from left to right, and (ii) have
the same binding power as $+$ and $-$ .

Operators $\maxplus/\maxminus$ are syntactic sugar for standard Presburger
terms:
we can rewrite a formula $\Gamma$ containing $x \maxplus[b] y$, using a
fresh variable $v$ per occurrence:
\begin{equation}
  \label{equation: maxplus rewriting}
  \Gamma \wbox{$\equiv$} (\Gamma|_{(x \maxplus[b] y) \rightarrow v}) \ \land \ ((x + y \geq b \land v = x + y) \lor (x + y < b \land v = b))
\end{equation}
where $\alpha|_{\beta \rightarrow \gamma}$ denotes substitution of $\gamma$
for $\beta$ in $\alpha$.

The \emph{summary} of a path $\quotient \sigma$ in $\ttdquotient$ for local
state $\ttdlocal$ that visits only trivial SCCs is
computed in \algorithmref{Symbolically executing a path for local state l}
by symbolically executing $\quotient \sigma$. The path is traversed
backwards; for certain edges a ``contribution'' to counter $n_{\ttdlocal}$
is recorded, namely for each edge of $\ttdtranssexpand$ that is adjacent to
local state $\ttdlocal$, but only if it is real, or it is an expansion edge
that starts in local state $\ttdlocal$. Note that the three \plkeyword{if}
clauses in \algorithmref{Symbolically executing a path for local state l}
are not disjoint:
the first two both apply when edge $e_i$ is ``vertical'': it both enters
and exits local state~$\ttdlocal$. In this case the two contributions
cancel out.
\begin{algorithm}[htbp]
  \begin{algorithmic}[1]
    \REQUIRE path $\quotient \sigma = \range[]{t_1}{t_k}$ in $\ttdquotient$, i.e. $(t_i,t_{i+1}) \in \ttdtranssexpand$ for $1 \atm i < k$ ; local state $\ttdlocal$
    \STMT \code{$e_i$ := $(t_i,t_{i+1})$ for $1 \atm i < k$ , $(s_i,l_i)$ := $t_i$ for $1 \atm i \atm k$}
    \STMT \code{summary := "$n_{\ttdlocal}$"} \COMMENT{\code{summary} is a string}
    \FOR {\code{$i$ := $k-1$} \plkeyword{downto} 1}
      \IF {$e_i \in \ttdtranss$ and $l_i = \ttdlocal$}
        \STMT \code{summary := summary."+1"} \COMMENT{\code{.} = string concatenation}
      \ENDIF
      \IF {$e_i \in \ttdtranss$ and $l_{i+1} = \ttdlocal$}
        \STMT \code{summary := summary."-1"}
      \ENDIF
      \IF {$e_i \in \ttdtranssexpand \setminus \ttdtranss$ and $l_i = \ttdlocal$}
        \STMT \code{summary := summary."$\maxminus$1+1"}
      \ENDIF
    \ENDFOR
    \RETURN \code{summary}
  \end{algorithmic}
  \caption{Exact path summary via symbolic execution}
  \label{algorithm: Symbolically executing a path for local state l}
\end{algorithm}

\begin{figure}[htbp]
  \centerTwoOut{%
    \begin{minipage}{.4\textwidth}
		\begin{tikzpicture}[>=stealth, scale=1.0]
    		\draw plot[mark=*,mark options={fill=white},xshift=0cm] file {};
    		\draw[->,xshift=0cm] (0,0) -- coordinate (x axis mid) (2.6,0) node[anchor=north]{$l$};
    		\draw[->,yshift=0cm] (0,0) -- coordinate (y axis mid)(0,2.6) node[anchor=east]{$s$};
    		\foreach \x/\xtext in {0.2/0,1.2/1,2.2/2}
        		\draw (\x cm,2.5pt) -- (\x cm,-3pt) node[anchor=north] {$\xtext$};
	    	\foreach \y/\ytext in {0.2/0,1.2/1, 2.2/2}
       			\draw (2.5pt,\y cm) -- (-3pt,\y cm) node[anchor=east] {$\ytext$};
    		\draw[->, thick] (0.2,0.2) node[right,font=\tiny]{$(0,0)$} -- (0.2,1.2)node[above,font=\tiny]{$(1,0)$};
    		\draw[->,dashed, thick] (0.2,1.2) -- (1.2,1.2)node[right,font=\tiny]{$(1,1)$};
    		\draw[->, thick] (1.2,1.2) -- (2.1,2.1)node[right,font=\tiny]{$(2,2)$};
		\end{tikzpicture}
    \end{minipage}}{%
    \begin{minipage}{.55\textwidth}
      Summary functions for local states $\ttdlocal=0,1,2$:
      \[
      \begin{array}{rclcl}
        \Sigma_0(n_0) & = & n_0 \maxminus 1 + 1 - 1 + 1 & = & n_0 \maxminus 1 + 1 \\
        \Sigma_1(n_1) & = & n_1 + 1 \\
        \Sigma_2(n_2) & = & n_2 - 1
      \end{array}
      \]
      Examples:
      \[
        \Sigma_0(0) = 1, \ \Sigma_0(1) = 1, \ \Sigma_1(0) = 1, \ \Sigma_2(1) = 0 \ .
      \]
    \end{minipage}}
  \caption{A quotient structure $\ttdquotient$ with a vertical edge}
  \label{figure: vertical edge ttd}
\end{figure}

The summary of path $\quotient \sigma$ for local state $\ttdlocal$ defines
a function $\func{\Sigma_{\ttdlocal}}{\NN}{\NN}$ that summarizes the effect
of path $\quotient \sigma$ on counter $n_{\ttdlocal}$. The summary
functions for the short path in \figureref{vertical edge ttd} are shown
next to the figure. These examples illustrate how we can encode a quotient
path that visits only trivial SCCs into a quantifier-free Presburger
formula. The formula for $\Sigma_0(n_0)$ implies that if we traverse the
path backwards from a state with $n_0=0$ threads in local state 0, at the
end there will be $\Sigma_0(0) = 0 \maxminus 1 + 1 = 1$ thread in local
state 0. If we start with $n_0 = 1$, we also end up with $n_0 = 1$. Note
that the path cannot be traversed backwards starting with $n_2 = 0$, since
its endpoint is thread state $\threadstate 2 2$.

What remains to be resolved is the handling of non-trivial SCCs
along~$\quotient \sigma$. Such SCCs are contractions of loops in the
expanded structure $\ttdexpand$, to the effect that paths in $\ttdexpand$
are no longer finite; their summaries cannot be obtained by symbolic
execution. Loops are of course the classical ``nuisance'' when expressing
reachability as a satisfiability problem. We address it in the rest of this
paper.

\section{Exact Acceleration of Simple Loops}
\label{section: Reachability as a Decision Problem: Simple Loops}

In this section we generalize path summaries to the case of a quotient path
$\quotient \sigma$ that visits SCCs formed by a \emph{single simple loop},
i.e., a single cyclic path without repeated inner nodes. In contrast to
unwinding approaches such as bounded model checking, we are aiming here at
an exact solution. Namely, for each loop~$\Loop$, we seek a closed form for
the value of local state counter $n_{\ttdlocal}$ after the backward search
\algorithmref{Abdulla} traverses $\Loop$ some number of times $\iterator$.

In this section, since we need to ``zoom in'' to SCCs collapsed into single
nodes in $\ttdquotient$, we instead look at paths in $\ttdexpand$. Recall
that for a straight-line path $\expand \sigma = \range[]{t_1}{t_k}$, the
value of counter $n_{\ttdlocal}$ after \algorithmref{Abdulla} traverses
$\expand \sigma$ can be computed using $\expand \sigma$'s path summary
function $\Sigma_{\ttdlocal}$, determined via symbolic execution
(\algorithmref{Symbolically executing a path for local state l}). We now
establish a lemma that renders the summary function suitable for
acceleration, in case the path is cyclic. As in \algorithmref{Symbolically
  executing a path for local state l}, we define $(s_i,l_i) := t_i$ for $1
\atm i \atm k$. Let
\begin{equation}
  \begin{array}{rclc}
    \delta_{\ttdlocal} & \ = \ & |\{i: 1 \atm i < k: \ (t_i,t_{i+1}) \in \ttdtranss \ \land \ \pos[l]{$l_i$}{$l_{i+1}$} = \ttdlocal\}| & \ - \\
                       &       & |\{i: 1 \atm i < k: \ (t_i,t_{i+1}) \in \ttdtranss \ \land \                 l_{i+1}   = \ttdlocal\}| &
  \end{array}
  \label{equation: delta}
\end{equation}
be the \emph{real-edge summary} $\delta_{\ttdlocal} \in \ZZ$ of $\expand
\sigma$, i.e.\ the number of \emph{real} edges along $\expand \sigma$ that
start in local state $\ttdlocal$, minus the number of \emph{real} edges
along $\expand \sigma$ that end in $\ttdlocal$. Value $\delta_{\ttdlocal}$
summarizes the total contribution by real edges to counter $n_{\ttdlocal}$
as path $\expand \sigma$ is traversed backwards: real edges starting in
$\ttdlocal$ increment the counter, those ending in $\ttdlocal$ decrement
it. The following lemma uses the $\delta_{\ttdlocal}$'s to compactly
determine local state $\ttdlocal$'s summary along $\expand \sigma$:
\newcounterset{lemmaCounterValuesSingleIteration}{\theDEF}
\begin{LEM}[proof in \appendixref{Proof of Lemma counter values single iteration}]
  \label{lemma: counter values single iteration}
  Let $b_{\ttdlocal} = \Sigma_{\ttdlocal}(1)$ if $l_k = \ttdlocal$ (path
  $\expand \sigma$ ends in local state $\ttdlocal$), and $b_{\ttdlocal} =
  \Sigma_{\ttdlocal}(0)$ otherwise. Then $\Sigma_{\ttdlocal}(n_{\ttdlocal})
  = n_{\ttdlocal} \maxplus[b_{\ttdlocal}] \delta_{\ttdlocal}$ .
\end{LEM}
The lemma suggests: in order to determine local state $\ttdlocal$'s summary
function in compact form, first compute the constant
$\Sigma_{\ttdlocal}(1)$ (or $\Sigma_{\ttdlocal}(0)$) using
\algorithmref{Symbolically executing a path for local state l}.
$\Sigma_{\ttdlocal}(n_{\ttdlocal})$ is then the formula as specified in the
lemma. The distinction whether path $\expand \sigma$ ends in state
$\ttdlocal$ is necessary intuitively because in this case the backward
traversal must start from a state with at least one thread in $\ttdlocal$.

Consider now a \emphasize{loop} $\expand \sigma =
\range[]{t_1}{t_{k-1}},t_1$ in $\ttdexpand$. Let $\Sigma_{\ttdlocal}$ be
$\expand \sigma$'s summary function for local state $\ttdlocal$, and
let $\delta_{\ttdlocal}$ and $b_{\ttdlocal}$ be defined as above in \equationref[]{delta} and
\lemmaref{counter values single iteration}, for the loop path $\expand
\sigma$.
\newcounterset{theoremCounterValues}{\theDEF}
\begin{THE}[proof in \appendixref{Proof of Theorem counter values}]
  \label{theorem: counter values}
  Let superscript $\iterate{}$ denote $\iterator$ function applications.
  Then, for $\iterator \atl 1$,
  \begin{equation}
    \label{equation: counter values}
    \iterate{\Sigma_{\ttdlocal}}(n_{\ttdlocal}) = \Sigma_{\ttdlocal}(n_{\ttdlocal}) \maxplus[b_{\ttdlocal}] (\iterator-1) \cdot \delta_{\ttdlocal} \ .
  \end{equation}
\end{THE}
In \equationref[]{counter values}, term $\Sigma_{\ttdlocal}(n_{\ttdlocal})$
marks the contribution to counter $n_{\ttdlocal}$ of the first loop
traversal, while $(\iterator-1) \cdot \delta_{\ttdlocal}$ marks the
contribution of the remaining $\iterator-1$ traversals.\footnote{By
  \lemmaref{counter values single iteration}, the right-hand side in
  \equationref[]{counter values} equals $n_{\ttdlocal}
  \maxplus[b_{\ttdlocal}] \delta_{\ttdlocal} \maxplus[b_{\ttdlocal}]
  (\iterator-1) \cdot \delta_{\ttdlocal}$, which turns out to be not quite
  equal to
    $n_{\ttdlocal} \maxplus[b_{\ttdlocal}] \iterator \cdot
  \delta_{\ttdlocal}$, due to the lack of associativity of $\maxplus$.
}

\paragraph{Example.}
We show how the reachability of thread state $\threadstate 6 4$ for the TTD
shown in \figureref{ttd abstractions} is analyzed symbolically. For each
local state $\ttdlocal \in \range 0 4$, the following constraints are
obtained by applying \lemmaref{counter values single iteration},
\theoremref{counter values}, and \lemmaref{counter values single iteration}
to the straight-line path from $\threadstate 6 4$ backwards to
$\threadstate 3 1$, the loop from $\threadstate 3 1$ to $\threadstate 3 1$,
and to the path from $\threadstate 3 1$ to
$\threadstate 0 0$, respectively:
\[
\begin{array}{lrrcrcrrrcrccr}
  n_0: \quad & 0 & \maxplus[0] &   0  & \maxplus[2] &   2  & \maxplus[2] & (k-1) & \cdot &   2  & \maxplus[3] &   3  & \mbox{} \quad \atl \quad \mbox{} & 1 \\
  n_1:       & 0 & \maxplus[1] &   0  & \maxplus[1] & (-1) & \maxplus[1] & (k-1) & \cdot & (-1) & \maxplus[0] & (-3) &         =                        & 0 \\
  n_2:       & 0 & \maxplus[2] &   2  & \maxplus[0] & (-1) & \maxplus[0] & (k-1) & \cdot & (-1) & \maxplus[0] &   0  &         =                        & 0 \\
  n_3:       & 0 & \maxplus[0] & (-2) & \maxplus[0] &   0  & \maxplus[0] & (k-1) & \cdot &   0  & \maxplus[0] &   0  &         =                        & 0 \\
  n_4:       & 1 & \maxplus[1] &   0  & \maxplus[0] &   0  & \maxplus[0] & (k-1) & \cdot &   0  & \maxplus[0] &   0  &         =                        & 0
\end{array}
\]
The equation for $n_4$ simplifies to $1 = 0$ and thus immediately yields
unsatisfiability and thus unreachability of the target thread state
$\threadstate 6 4$. In contrast, for target thread state $\threadstate 6
3$, the equations for $n_3$ and $n_4$ both reduce to $\true$. The
conjunction of all five equations reduces to $1 \maxplus[0] (\iterator-1)
\cdot (-1) = 0$. This formula is satisfied by $\iterator=2$, claiming
reachability of $\threadstate 6 3$ via a path containing two full
iterations of the loop from $\threadstate 3 1$ to $\threadstate 3 1$. Since
our method is exact for the case of simple loops, this path is guaranteed
to be genuine.

\section{Summarizing Loop Nests using Recurrence Relations}
\label{section: Strongly Connected Components with Loop Nests}

Suppose an SCC along a quotient path $\quotient \sigma$ contains several
simple loops, i.e.\ a ``loop nest'', such as in the ETTD shown in
\figureref{an example of TTD with loop nests} (left). Loops $\Loop_1$ and
$\Loop_2$ permit many structurally different paths, for instance those of
the form $\kleene{(\kleene{\Loop_1} \concat \Loop_2)}$.
The part $\kleene{\Loop_1} \concat \Loop_2$
does not correspond to a fixed straight-line path;
\theoremref{counter values} can thus not be applied to accelerate the outer
loop.
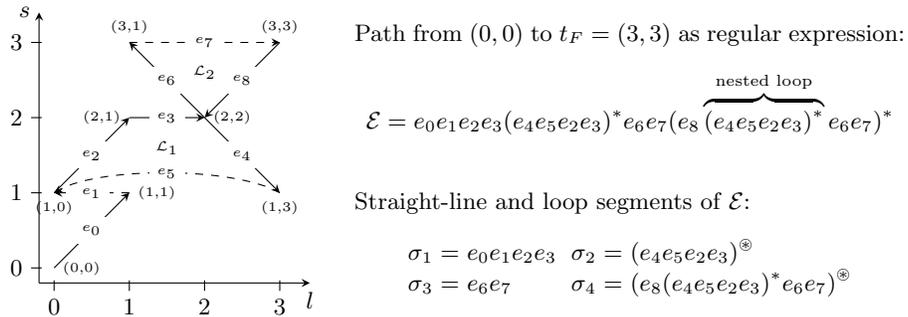
\begin{figure}[htbp]
  \centerTwoOut{%
    \begin{minipage}{0.38\textwidth}
      \centering
      \begin{tikzpicture}
        \begin{scope}[only marks, >=stealth]
          \draw plot[mark=*,mark options={fill=white},xshift=0cm] file {};
          \draw[->,xshift=0cm] (0,0) -- coordinate (x axis mid) (3.6,0) node[anchor=north]{$l$};
          \draw[->,yshift=0cm] (0,0) -- coordinate (y axis mid) (0,3.6) node[anchor=east]{$s$};
          \foreach \x/\xtext in {0.2/0,1.2/1,2.2/2,3.2/3}
          \draw (\x cm,1pt) -- (\x cm,-3pt) node[anchor=north] {$\xtext$};
          \foreach \y/\ytext in {0.2/0,1.2/1, 2.2/2,3.2/3}
          \draw (1pt,\y cm) -- (-3pt,\y cm) node[anchor=east] {$\ytext$};

          \draw[->] (0.2,0.2) node[right,font=\tiny]{(0,0)} -- (1.2,1.2) node[right,font=\tiny]{(1,1)} node [midway, fill=white, font=\tiny] {$e_0$};
          \draw[->,dashed] (1.2,1.2) -- (0.2,1.2) node[below,font=\tiny]{(1,0)} node [midway, fill=white, font=\tiny] {$e_1$};
          \draw[->] (1.2,2.2) -- (2.2,2.2) node[right,font=\tiny]{(2,2)} node [midway, fill=white, font=\tiny] {$e_3$};
          \draw[->] (0.2,1.2) -- (1.2,2.2) node[left,font=\tiny]{(2,1)} node [midway, fill=white, font=\tiny] {$e_2$};
          \draw[->] (2.2,2.2) -- (3.2,1.2) node[below,font=\tiny]{(1,3)} node [midway, fill=white, font=\tiny] {$e_4$};
          \draw[->] (2.2,2.2) -- (1.2,3.2) node[above,font=\tiny]{(3,1)} node [midway, fill=white, font=\tiny] {$e_6$};
          \draw[->, dashed] (1.2,3.2) -- (3.2,3.2)node[above,font=\tiny]{(3,3)} node [midway, fill=white, font=\tiny] {$e_7$};
          \draw[->] (3.2,3.2) -- (2.2,2.2) node [midway, fill=white, font=\tiny] {$e_8$};
          \draw[->, dashed] (3.2,1.2) .. controls +(150:0.7cm) and +(30:0.7cm) .. (0.2,1.2) node [midway, fill=white, font=\tiny] {$e_5$};
          \node (l1) [xshift=1.7cm, yshift=1.8cm, font=\tiny] {$\Loop_1$};
          \node (l2) [xshift=2.2cm, yshift=2.8cm, font=\tiny] {$\Loop_2$};
        \end{scope}
      \end{tikzpicture}
    \end{minipage}}{
    \begin{minipage}{0.60\textwidth}
      Path from $\threadstate 0 0$ to $\ttdfinalstate = \threadstate 3 3$
      as regular expression:
      \[
      \RE=e_0e_1e_2e_3\kleene{(e_4e_5e_2e_3)}e_6e_7
           {(e_8\overbrace{\kleene{(e_4e_5e_2e_3)}}^{\text{nested loop}}e_6e_7)}{}^*
      \]

      \

      Straight-line and loop segments of $\RE$:
      \[
      \begin{array}{rclcrcl}
        \sigma_1 & = & e_0e_1e_2e_3 &\qquad & \sigma_2 & = & \iterate[\outermostmark]{(e_4e_5e_2e_3)} \\
        \sigma_3 & = & e_6e_7       & & \sigma_4 & = & \iterate[\outermostmark]{(e_8\kleene{(e_4e_5e_2e_3)}e_6e_7)}
      \end{array}
      \]
    \end{minipage}}
  \caption{An ETTD with a loop nest (left); its regular expression decomposition (right)}
  \label{figure: an example of TTD with loop nests}
\end{figure}

Our approach to handling complex loops, inspired in part by
\cite{Farzan2015}, is to overapproximate their behavior in the form of a
\emph{transition invariant}. We first capture the set of paths $\expand \sigma$
in $\ttdexpand$ from $\ttdinitstate$ to $\ttdfinalstate$ represented by
quotient path $\quotient \sigma$ as a regular expression~$\RE$ (see
\figureref{an example of TTD with loop nests}, top right). We use a
standard algorithm \cite{Brzozowski:1964:DRE:321239.321249} to unravel the
loop structure inside non-trivial SCCs. The resulting expression $\RE$ can
be written using only concatenation and Kleene star $\kleene{}$: since
$\quotient \sigma$ is an SCC quotient path, the choice operator $\choice$
occurs in the translation at most inside loops and can thus be eliminated
using the identity $\kleene{(\RETWO \choice \RETHREE)} =
\kleene{(\kleene\RETWO \concat \kleene\RETHREE)}$ (see
\appendixref{Making Regular Expressions Alternation-Free}).

We next identify straight-line and loop segments in $\RE$, as shown in
\figureref{an example of TTD with loop nests}, bottom right. Each
straight-line segment is summarized exactly in a Presburger formula via
\lemmaref{counter values single iteration}. Loops $\Loop$ are processed
recursively as follows. If $\Loop = \kleene \re$ is \emph{innermost}
(i.e.\ $\re$ is straight-line), it is accelerated exactly using
\theoremref{counter values}, resulting in a Presburger formula of the form
$\iterate{\Sigma_{\ttdlocal}}(n_{\ttdlocal})$, for local state~$\ttdlocal$
and loop iterator $\iterator$. If $\Loop$ is not innermost, we first
recursively translate expression $\re$ into a transition invariant
$\invariant$ over $n_{\ttdlocal}$ and $n_{\ttdlocal}'$ and then solve the
\emph{recurrence relation} $\iterate{\invariant}$ ($\iterator$-fold
application). This step is described in more detailed below.

Orthogonally to ``innermost'', we differentiate whether $\Loop$ is
\emph{outermost}, or itself \emph{nested} in another loop above it. In the
latter case, we need to
summarize the behavior of $\Loop$ independently of the number $\iterator$
of iterations.
This is achieved by existentially abstracting $\iterator$ from the summary
formula, followed by standard Presburger quantifier elimination. Finally,
$\Loop$ may be outermost; such loops are marked in $\RE$ by the iterator
symbol~$\outermostmark$,
e.g.\ $\Loop = \iterate[\outermostmark] \re$. For such loops, parameter
$\iterator$ is retained: it becomes a free variable in the final Presburger
formula; a satisfying assignment, if any, specifies the number of
iterations of this outermost loop.

\Paragraph

This procedure is formalized in \algorithmref{path summary via invariant}.
Input is the regular expression $\RE$ obtained from path $\quotient
\sigma$, split into straight-line and loop segments
$\range[]{\sigma_1}{\sigma_p}$, and a local state $\ttdlocal$.
The algorithm walks through $\RE$ backwards (\lineref{for loop beginning}),
processing straight-line segments (\lineref{linear path}) and four types of
loops depending on what combination of ``outermost (\omost)'' and
``innermost (\imost)'' they fall in. The transition invariants $\invariant$
for the individual segments are composed via relational product, denoted
$\compos$. Output is a
Presburger formula $\invariant$ that summarizes the effect of any path
$\expand \sigma$ represented by $\quotient \sigma$, as a transition
invariant over local state $\ttdlocal$'s counter variable $n_{\ttdlocal}$,
its post-path value~$n_{\ttdlocal}'$, and the iterators $\iterator_i$ for
the outermost loops.
\begin{algorithm}[htbp]
  \begin{algorithmic}[1]
    \REQUIRE regular expression $\RE = \sigma_1 \concat \ldots \concat \sigma_p$ with $q$ outermost loops; local state $\ttdlocal$ 
    \ENSURE $\invariant(n_{\ttdlocal},n_{\ttdlocal}',\range[]{\iterator_1}{\iterator_q})$: a Presburger summary for $\RE$ and local state $\ttdlocal$
    \STMT $\invariant := \true$  \label{line: initialize phi}
    \FOREACH[symbolically execute $\quotient \sigma$ backwards]{$i = p\ \downto\ 1$}                                              \label{line: for loop beginning}
      \IF {$\sigma_i$ is straight-line}
        \STMT $\invariant := \invariant \compos \Sigma_{\ttdlocal}(n_{\ttdlocal})$
        \COMMENT{$\Sigma_{\ttdlocal}(n_{\ttdlocal})$: \lemmaref{counter values single iteration}}                                 \label{line: linear path}
      \ELSIF{$\sigma_i$ has the form $\kleene\re$}
        \IF{$\re$ is star-free}
          \STMT $\invariant := \invariant \compos \exists\iterator.\ \iterate{\Sigma_{\ttdlocal}}(n_{\ttdlocal})$
          \COMMENT{$\sigma_i$ not \omost\ but \imost. $\iterate{\Sigma_{\ttdlocal}}(n_{\ttdlocal})$: \theoremref{counter values}} \label{line: not om, im}
        \ELSE
          \STMT $\invariant_i := \pathformula(\re,\ttdlocal)$ \COMMENT{$\sigma_i$ not \omost, not \imost}                         \label{line: not om, not im 1}
          \STMT  $\invariant_i(\iterator) := \textsc{Solve-Recurrence}(\iterate{\invariant_i})$                                   \label{line: not om, not im 2}
          \STMT  $\invariant := \invariant \compos \exists\iterator.\ \invariant_i(\iterator)$                                    \label{line: not om, not im 3}
        \ENDIF
      \ELSIF{$\sigma_i$ has the form $\iterate[\outermostmark]\re$}
        \IF{$\re$ is star-free}
          \STMT $\invariant := \invariant \compos \iterate{\Sigma_{\ttdlocal}}(n_{\ttdlocal})$
          \COMMENT{$\sigma_i$ \omost\ and \imost. $\iterate{\Sigma_{\ttdlocal}}(n_{\ttdlocal})$: \theoremref{counter values}}     \label{line: om, im}
        \ELSE
          \STMT $\invariant_i := \pathformula(\re,\ttdlocal)$ \COMMENT{$\sigma_i$ \omost\ but not \imost}                         \label{line: om, not im 1}
          \STMT $\invariant := \invariant \compos \textsc{Solve-Recurrence}(\iterate{\invariant_i})$                              \label{line: om, not im 2}
        \ENDIF
      \ENDIF
    \ENDFOR
    \RETURN $\ttdlocal = \ttdinitlocal$ ? $\invariant \atl 1$ : $\invariant = 0$
  \end{algorithmic}
  \caption{$\pathformula(\RE,\ttdlocal)$}
  \label{algorithm: path summary via invariant}
\end{algorithm}

\paragraph{Acceleration via recurrence solving.} If expression $\re$ in
$\sigma_i = \kleene\re$ or $\sigma_i = \iterate[\outermostmark]\re$
contains loops on its own, we summarize $\sigma_i$ by closing the
transition relation invariant obtained for $\re$ under $\iterator$-fold
recurrence. Solving such recurrences
turns out to be manageable, as all involved formulas are in linear integer
arithmetic extended by the $\maxplus$ operator. We rewrite the $\maxplus$
according to \equationref{maxplus rewriting} and convert the resulting
formula into disjunctive normal form. We search each disjunct separately
for a solution.
For each disjunct we push the recurrence operator $\iterate{}$ inside and
apply it only to individual conjuncts; \appendixref{Recurrences of
  Conjunctions as Conjunctions of Recurrences} justifies.

Each conjunct is of the elementary form $n' \relation c$, $n \relation c$,
or $n' \relation n + c$, where $\relation\ \in \{\atm, =, \atl\}$ and $c
\in \ZZ$. We solve the $\iterator$-fold recurrence of (i.e.,
``accelerate'') these elementary relations as follows. Relations $n'
\relation c$ and $n \relation c$ are invariant under $\iterator$-fold
acceleration. Relation $n' \relation n + c$ is accelerated according to the
\linebreak

\vspace{-8pt}

\noindent
\centerTwoOut{%
  \begin{minipage}{47mm}
    table on the right. Here, $n$ is the variable value at path entry,
    $\iterate n$ the value after $\iterator$-fold acceleration, and $n'$
    the value after abstracting the number $\iterator$ of loop iterations.
  \end{minipage}}{%
  \begin{minipage}{70mm}
    \resizebox{7cm}{!}{
      \raggedright
      \begin{tabular}{|c||c||c|c|c|}
        \hline
        \multirow{2}{*}{$\relation$} & \emphasize{$\iterator$-fold accelerati-} & \multicolumn 3 {c|} {\emphasize{``$\exists \iterator \ldots$'' + quantif.\ elimin.}} \\
        \cline{3-5}
                                     & \emphasize{on of $n' \relation n + c$} & $c > 0$ & $c = 0$ & $c < 0$ \\
        \hline
        $\atl$ & $\iterate n \atl n + \iterator \cdot c$ & $n' \atl n + c$ & $n' \atl n$ & $\true$ \\
        $\atm$ & $\iterate n \atm n + \iterator \cdot c$ & $true$          & $n' \atm n$ & $n' \atm n + c$ \\
        $=$    & $\iterate n  =   n + \iterator \cdot c$ & $n' \atl n + c$ & $n' = n$    & $n' \atm n + c$\\
        \hline
    \end{tabular}}
  \end{minipage}}

\vspace{3pt}

\paragraph{Example.} We revisit \figureref{an example of TTD with loop
  nests}. Given the regular expression $\RE$ shown in the figure,
\algorithmref{path summary via invariant} constructs the following
constraints for the four local states ($\iterator_1$ is the loop iterator
for $\sigma_2$, $\iterator_2$ is that for the outer loop of segment
$\sigma_4$):
\begin{equation*}
  \begin{array}{lclclcl}
     l = 0: & & n'_0 \atl 1 \land n'_0  =  \iterator_2 + \iterator_1 + 2 &               & l = 1: & & n'_1 = 0 \\
     l = 2: & & n'_2 = 0                                                 & \hspace*{5mm} & l = 3: & & 0 \atm n'_3 \atm 1 -\iterator_2 \lor n'_3 = 0
  \end{array}
  \label{equation: example: loop nests after simplification}
\end{equation*}
The conjunction of these four constraints is satisfiable; a solution is
$\iterator_1 = 1$ and $\iterator_2 = 0$.
We cannot, however, conclude that $\ttdfinalstate$ is reachable in
$\ttdinf$: the solution may be spurious, as we have overapproximated the
nested-loop behavior. We finally therefore design an algorithm that tries
to settle this
question.

\paragraph{Path reachability.} Given a quotient path $\quotient \sigma$ and
a corresponding regular expression $\RE$, \algorithmref{reachability}
attempts to decide the reachability of final thread state $\ttdfinalstate$
in $\ttdinf$ along a path represented by $\quotient \sigma$. If, for each
local state $\ttdlocal$, the Presburger formula determined by
\algorithmref{path summary via invariant} is unsatisfiable,
$\ttdfinalstate$ is unreachable along $\quotient \sigma$. Otherwise we have
a satisfying assignment $\iterator_i$ to the iterators for the outermost
loops in $\RE$ (those for nested loops have been abstracted away).
We now unwind each outermost loop $\Loop_i$ in $\RE$ $\iterator_i$ times
--- we think of this as ``peeling away'' the outermost loop layer. As a
result, the loop nesting height in $\RE$ decreases by one. We repeat the
satisfiability question from above for each local state.

This process has two possible outcomes: if any iteration of the
\plkeyword{while} loop in \lineref{big while} yields \emph{unsat}
(\lineref{satisfiability check 2}), we return \emph{unknown along
  $\quotient \sigma$}: at this point formula $\pathformula(\RE,\ttdlocal)$
no longer overapproximates, due to the partially instantiated
loop iterators.
Otherwise, since the nesting height decreases in each iteration, $\RE$ will
eventually be loop nest free. The iterator assignment $\{\iterator_i\}$ is
now complete and can be unwound into a linear a path, which is checked for
genuineness (Lines~\lineref[]{unwind 2}--\lineref[]{unknown 2}).
\begin{algorithm}[htbp]
  \begin{algorithmic}[1]
    \REQUIRE $\RE$: regular expression for quotient path $\quotient \sigma$
    \ENSURE $\{$ \emph{unreachable} \ $\choice$ \ \emph{reachable} + witness path \ $\choice$ \ \emph{unknown} $\}$ \emph{along $\quotient \sigma$} 
    \IF{$\LAND_{\ttdlocal \in \ttdlocals} \pathformula(\RE,\ttdlocal)$ \ is
      unsatisfiable}                            \label{line: satisfiability check 1}
      \RETURN \emph{unreachable along $\quotient \sigma$}                                                             \label{line: unreachable}
    \ENDIF
    \STMT $\range[]{\iterator_1}{\iterator_q} := \mbox{satisfying assignment}$                                        \COMMENT{$q$: current \# of outermost loops}
    \WHILE {$\RE$ contains loop nests}                                                                                \label{line: big while}
      \STMT $\RE := \mbox{\textsc{Unwind}}(\RE, \range[]{\iterator_1}{\iterator_q})$                                  \label{line: unwind 1}
      \IF{$\LAND_{l \in \ttdlocals}\pathformula(\RE, l)$ \ is unsatisfiable}                                          \label{line: satisfiability check 2}
        \RETURN \emph{unknown along $\quotient \sigma$}                                                               \label{line: unknown 1}
      \ENDIF
      \STMT $\range[]{\iterator_1}{\iterator_q} := \mbox{satisfying assignment}$                                      \COMMENT{$q$: current \# of outermost loops}
    \ENDWHILE
    \IF{$\mbox{\textsc{Unwind}}(\RE, \range[]{\iterator_1}{\iterator_q})$ represents a feasible execution path}       \label{line: unwind 2}
      \RETURN \emph{reachable} + witness path                                                                         \label{line: reachable}
    \ELSE
      \RETURN \emph{unknown along $\quotient \sigma$}                                                                 \label{line: unknown 2}
    \ENDIF
  \end{algorithmic}
  \caption{$\mbox{Path-Reachability}(\RE)$}
  \label{algorithm: reachability}
\end{algorithm}

\paragraph{Example (continued).} \algorithmref{reachability} confirms that
the assignment $\iterator_1 = 1,\iterator_2 = 0$ found above for the scene
in \figureref{an example of TTD with loop nests} corresponds to a genuine
path, given by the edge sequence $e_0e_1e_2e_3e_4e_5e_2e_3e_6e_7$. This
proves $\ttdfinalstate$ reachable.


\newcounterset{theoremSoundness}{\theDEF}
\begin{THE}[Soundness; proof in \appendixref{Proof of Theorem soundness}]
  \label{theorem: soundness}
  If, for each quotient path $\quotient \sigma$ from $\ttdinitstate$'s to
  $\ttdfinalstate$'s SCC, \algorithmref{reachability} returns
  \emph{unreachable}, then $\ttdfinalstate$ is unreachable in $\ttdinf$.
\end{THE}
Termination of \algorithmref{reachability} is guaranteed since the loop
nesting height decreases in each iteration of the \plkeyword{while} in
\lineref{big while}. Moreover, as \appendixref{Correctness for the
  Simple-Loop Case} shows: if $\RE$ is loop nest free,
\algorithmref{reachability} never returns unknown.
This is in particular the case when all loops in $\ttdexpand$ are simple;
the algorithm is sound and complete for such systems.

\section{Empirical Evaluation}
\label{section: Implementation}

Our technique is implemented in a reachability checker named \ourtool\ (for
``Un\-bounded-thread Reachability via Symbolic execUtion and Loop
Acceleration'').

\format{\vspace{0em}}
\subsubsection{Benchmarks and Experimental Setup.}
We evaluate our technique on a collection of 60 examples, which is
organized into two suites. The first suite contains 30 Petri nets (taken
from~\cite{ELMMN14}), 26 of which are safe. The second suite contains 30
Boolean programs generated from \C\ programs (taken from~\cite{KKW12})
using \SATABS, 5 of which are safe.
For each benchmark, we consider verification of a reachability property. In
the case of \C\ programs, the property is specified via an assertion. We
excluded some benchmarks from~\cite{ELMMN14,KKW12}, because they have
certain features (e.g.\ broadcast transitions) that \ourtool\ currently
does not support.

To apply \ourtool\ to \C\ programs, we use \SATABS\ to transform those
programs to TTDs (option \texttt{--build-tts}) via intermediate Boolean
programs~\cite{DKKW2011}. When \SATABS\ requires several CEGAR iterations
over the \C\ programs until the abstraction permits a decision, the same
\C\ source program gives rise to several Boolean programs and TTDs.
We use Z3 (v4.3.2) as the Presburger solver~\cite{MB08}. All experiments
are performed on a 2.3GHz Intel Xeon machine with 64 GB memory, running
64-bit Linux. Execution time is limited to 10000 seconds; memory to 4 GB.
All benchmarks and our tool are available
online~\cite{LoopAnalysisWebsite}.

Our evaluation is carried out in three steps: a comparison of
\ourtool\ against a recent constraint-based (``symbolic'') coverability
checker~\cite{ELMMN14}, against a range of traditional state space
exploration based coverability checkers, and against \BFC\ with and without
a \emph{coverability oracle} \cite{KKW12}.

\subsubsection{Comparison.} \figureref{comparison vs
  petrinizer and mcov-gkm} (left) plots the comparison against
\Petrinizer\footnote{\Petrinizer\ offers four methods; we use the most
  powerful -- refinement over integer}~\cite{ELMMN14}, a recent constraint
based coverability checker for Petri nets. It employs the \emph{marking
  equation} technique, which essentially considers unordered collections of
transitions, instead of firing sequences.
\begin{figure}[htbp]
  \begin{minipage}{0.32\linewidth}
    \begin{tikzpicture}[scale=0.50]
      \begin{loglogaxis}[
        height=7.5cm,
        width=7.5cm,
        xmin=0.001,xmax=10000,
        ymin=0.001,ymax=10000,
        xticklabels={,,,,$10^{0}$,$10^{1}$,$10^{2}$,$10^{3}$,$10^{4}$},
        xlabel={ \ourtool\ (sec.)},
        yticklabels={,,,,$10^{0}$,$10^{1}$,$10^{2}$,$10^{3}$,$10^{4}$},
        ylabel={ \Petrinizer\ (sec.)},]
        \addplot[only marks, mark=+, color=red] table {data/ourtool-petrinizer-PNs.dat};
        \draw[red] (axis cs:0.001,0.001) -- (axis cs:10000,10000);
        \addplot[only marks, mark=x, color=blue] table {data/ourtool-petrinizer-BPs.dat};
      \end{loglogaxis}
    \end{tikzpicture}
  \end{minipage}
  \hfill{%
  \begin{minipage}{0.32\linewidth}
    \begin{tikzpicture}[scale=0.50]
      \begin{loglogaxis}[
        height=7.5cm,
        width=7.5cm,
        xmin=0.001,xmax=10000,
        ymin=0.001,ymax=10000,
        xticklabels={,,,,$10^{0}$,$10^{1}$,$10^{2}$,$10^{3}$,$10^{4}$},
        xlabel={ \ourtool\ (sec.)},
        yticklabels={,,,,$10^{0}$,$10^{1}$,$10^{2}$,$10^{3}$,$10^{4}$},
        ylabel={ \BFC\ (sec.)},]
        \addplot[only marks, mark=x, color=blue] table {data/ourtool-mcov.dat};
        \draw[red] (axis cs:0.001,0.001) -- (axis cs:10000,10000);
      \end{loglogaxis}
    \end{tikzpicture}
  \end{minipage}}
  \hfill{%
  \begin{minipage}{0.32\linewidth}
    \begin{tikzpicture}[scale=0.50]
      \begin{loglogaxis}[
        height=7.5cm,
        width=7.5cm,
        xmin=0.001,xmax=10000,
        ymin=0.001,ymax=10000,
        xticklabels={,,,,$10^{0}$,$10^{1}$,$10^{2}$,$10^{3}$,$10^{4}$},
        xlabel={ \ourtool+\GKM\ (sec.)},
        yticklabels={,,,,$10^{0}$,$10^{1}$,$10^{2}$,$10^{3}$,$10^{4}$},
        ylabel={\BFC+\GKM\ (sec.)}]
        \addplot[only marks,mark=x,color=blue] table {data/ourtoolgkm-mcovgkm.dat};
        \draw[red] (axis cs:0.001,0.001) -- (axis cs:10000,10000);
      \end{loglogaxis}
    \end{tikzpicture}
  \end{minipage}}
  \caption{Comparison: left: \ourtool\ against \Petrinizer\ (for
    \Petrinizer, we mark ``unknown'' as timeout); center: \ourtool\ against
    \BFC; right: \ourtool+\GKM\ with \BFC+\GKM.\ Suffix ``+\GKM'' means
    tool uses Karp-Miller procedure as forward accelerator. Each dot
    represents execution time on one example. }
  \label{figure: comparison vs petrinizer and mcov-gkm}
\end{figure}
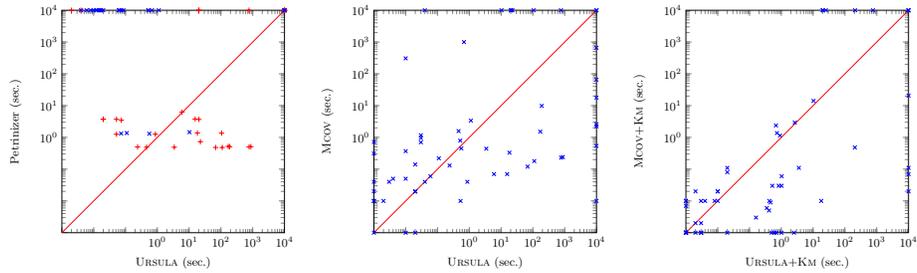

\noindent
\centerTwoOut{%
  \begin{minipage}{47mm}
    The table on the right classifies how many instances \Petrinizer\ and
    \ourtool\ can solve in each benchmark category. We note that
    \Petrinizer\ quickly discharges most \emph{safe} instances.
  \end{minipage}}{%
  \begin{minipage}{70mm}
    \resizebox{7cm}{!}{
      \raggedright
      \begin{tabular}{|l||c|c|c|c|}
        \hline
         \multicolumn{1}{|c||}{suite} & safe PN & unsafe PN & safe BP & unsafe BP\\
        \hline
        \Petrinizer\ & $22/26$\footnote{\# of proved instances/\# of
          total instances}  &  $0/4$     & $5/5$    & $\ \ 0/25$ \\
        \hline
        \ourtool\    & $24/26\ $  &  $2/4$     & $5/5$    & $20/25$ \\
        \hline
      \end{tabular}}
  \end{minipage}}
\ourtool\ is (much) more precise but, as \figureref{comparison vs
  petrinizer and mcov-gkm} shows, takes slightly more time.

For $1 \atm k \atm 60$, \figureref{time to solve k programs} plots
the total time (log-scale) taken to solve the $k$ easiest of our
benchmark problems, for the following tools:\\
\begin{tabular}{lll}
  \BWS:  & & Backward reachability analysis~\cite{Abdulla2010,ACJT96}
  (\algorithmref{Abdulla})\\
  \KM:   & & A Karp-Miller procedure~\cite{KKW12} (v1.0)\\
  \IIC:  & & Incremental, inductive coverability algorithm~\cite{KMNP13}\\
  \Mist-\textsc{Ar}: & & An abstraction refinement method presented in~\cite{GRV08} (v1.0.3)\\
\end{tabular}


The results in the plot demonstrate that \ourtool\
solves the most
benchmarks (51). 
\IIC\ is the most competitive among the other tools until the
benchmarks are reached that it cannot solve. In general, we observe
that other tools outperform \ourtool\ on small benchmarks, an effect
that can be explained by the overhead of path-wise analysis, regular
expression conversion and Z3. For instance, the percentage of
execution time spent on regular expression conversion is over
$50\%$ on average. How to effectively
build regular expressions for TTDs or Petri nets is a question left
for future research.
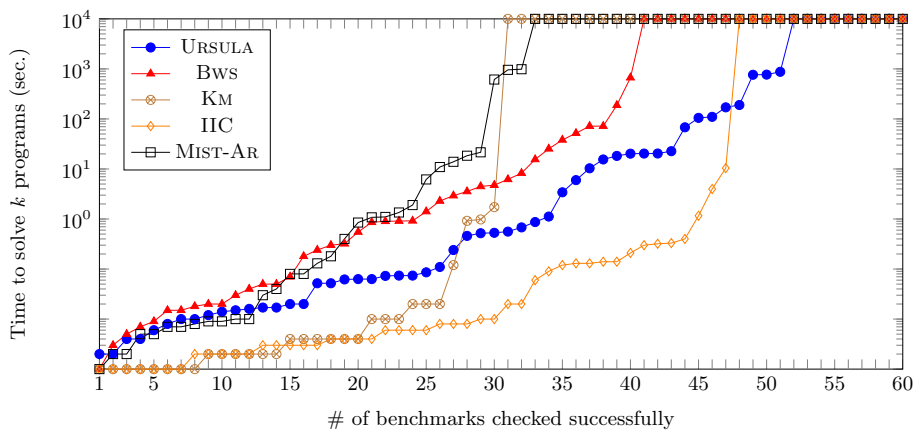
\begin{figure}[htbp]
  \centering
  \begin{tikzpicture}[scale=0.86]
    \begin{semilogyaxis}[
      height=7cm,
      width=14cm,
      legend pos=north west, 
      legend style={font=\small},
      xmin=1,xmax=60,
      ymin=0.001,ymax=10000,
      xtick={1,5,10,15,20,25,30,35,40,45,50,55,60},
      extra x ticks={1,2,3,4,6,7,8,9,11,12,13,14,16,17,18,19,21,22,23,24,26,27,28,29,31,32,33,34,36,37,38,39,41,42,43,44,46,47,48,49,51,52,53,54,56,57,58,59},
      extra x tick style={xticklabel=\empty},
      yticklabels={,,,,$10^{0}$,$10^{1}$,$10^{2}$,$10^{3}$,$10^{4}$},
      xlabel={$\#$ of benchmarks checked successfully},
      ylabel={Time to solve \emph{k} programs (sec.)}]
      \addplot [mark=*, color=blue] file {data/ourtool.dat};
      \addplot [mark=triangle*, color=red] file {data/bws.dat};
      \addplot [mark=otimes, color=brown] file {data/bfckm.dat};
      \addplot [mark=diamond,color=orange] file {data/iic.dat};
      \addplot [mark=square, color=black] file {data/mist.dat};
      \legend{\ourtool,\BWS,
        \KM,\IIC,\Mist-\textsc{Ar}}
    \end{semilogyaxis}
  \end{tikzpicture}
  \caption{Comparison: cactus plot comparing \ourtool\ with prior
    coverability tools. An entry of the form $(k,t)$ for some curve shows
    the time $t$ it took to solve the $k$ easiest --- for the method
    associate with that curve --- benchmarks (order varies across
    methods).}
  \label{figure: time to solve k programs}
\end{figure}

The center part of \figureref{comparison vs petrinizer and mcov-gkm} plots
the comparison against \BFC, a very efficient explicit-state exploration
method. \ourtool\ remains competitive, despite its relatively prototypical
character, and the comparatively long efforts that have gone into the
design of \BFC.
To investigate how our
technique fares against other \emph{backward-directed} techniques but
equipped with \emph{forward accelerators} (suggested first in
\cite{KKW12}), we pair \ourtool\ and \BFC\ with the Karp-Miller procedure.
The right part of \figureref{comparison vs petrinizer and mcov-gkm} plots
the comparison of execution time. We note that \BFC\ with KM performs
better -- it solves more instances faster -- than \ourtool. The difference
is explained by the tight and sophisticated integration of KM into \BFC,
whereas \ourtool\ is not able to benefit from forward reachability
information reported for non-query elements.
A~deeper integration of a forward accelerator into our algorithm is an
extension left for future work.

\section{Related Work}

Groundbreaking results in
infinite-state system analysis include the
decidability of coverability in \emph{vector addition systems} (VAS)
\cite{Karp1969}, and the work by German and Sistla on modeling
communicating finite-state threads as VAS~\cite{German1992}. Numerous
results have since
improved on the original
procedure in \cite{Karp1969} in practice \cite{GRB06,GRB07,RS11,VH12}.
Others extend it to more general computational models, including
\emph{well-structured} \cite{FS01} or \emph{well-quasiordered} (wqo)
transition systems \cite{ACJT96,Abdulla2010}.

The wqo-based approach, in basic form shown in \algorithmref{Abdulla},
along with work on \emph{acceleration} techniques for infinite-state
systems \cite{Finkel2002,JS07}, was inspirational for this paper: part of
our algorithm builds a Presburger formula while symbolically executing the
backward search process in \cite{Abdulla2010}.
Our treatment of complicated nested loop structures was inspired in part by
the
work in \cite{Farzan2015} on computing numerical transition invariants via
recurrence analyses.

Recent theoretical work by Leroux employs Presburger arithmetic to solve
the VAS global configuration reachability (not coverability) problem. In
\cite{Leroux2009}, it is shown that a state is \emph{unreachable} in the
VAS iff there exists an ``inductive`` Presburger formula that separates the
initial and final states. The existence of such a formula is determined by
enumeration; termination is guaranteed by running a second semi-algorithm
whose termination is guaranteed in the case of reachability. The
theoretical complexity of this technique is mostly left open. Practicality
is not discussed and doubted later by the author in \cite{Leroux2013},
where a more direct approach is presented that permits the computation of a
Presburger definition of the reachability set of the VAS in some cases,
e.g.\ for \emph{flatable} VAS. Reachability can then be cast as a
Presburger decision problem, as in our algorithm. The question under what
exact conditions the VAS reachability set is Presburger-definable appears
to be undecided.

The results referenced above are mainly foundational in nature and target
generally harder
reachability questions than we do in this paper. Our contribution here is
not to reproduce these theoretical results.
Instead, it is to show how to practically compute a Presburger encoding
whose unsatisfiability implies safety, and that the resulting formulas are
often very short and easy to decide, thus giving rise to an efficient
algorithm.

In recent work, classical techniques based on Petri net \emph{marking
  equations}
are revisited and used to reduce the coverability problem to linear
constraint solving~\cite{ELMMN14}.
Like our work, this approach
benefits from advances in SMT technology but is generally incomplete (the
constraints overapproximate coverability). We have
shown our symbolic encoding
to be (more complex and) more precise: our inputs are not generic Petri
nets, but systems derived from programs with shared state synchronization
that imposes partial control flow constraints.
Moreover, we have shown how to detect spuriousness of
solution paths at least in some cases; this issue is not addressed in
\cite{ELMMN14}.



\bibliographystyle{splncs03}
\bibliography{references}

\newpage

\appendix

\section*{Appendix}

\section{Uniqueness of the Initial State}
\label{appendix: Uniqueness of the Initial State}

It is inexpensive to enforce a unique initial thread state without
affecting thread state reachability, provided the initial thread state set
$T$ of the given TTD $\ttd$ satisfies the following ``box'' property:
\begin{equation}
  \forall (s,t) \in T, \ (s',t') \in T: \ (s,t') \in T, \ (s',t) \in T \ .
  \label{equation: box property}
\end{equation}
This holds if $T$ is a singleton. More generally, it holds if all states in
$T$ have the same shared state, and it holds if all states in $T$ have the
same local state. It also holds of a set $T$ whose elements form a complete
rectangle in the graphical representation of $\ttd$.

To enforce a unique initial thread state, we build a new TTD $\ttd'$ that
is identical to $\ttd$, except that it has a single initial thread state
$\ttdinitstate = \ttdinitstatepair$ with fresh shared and local states
$\ttdinitshared,\ttdinitlocal$, and the following additional edges:
\begin{eqnarray}
  \ttdinitstatepair         \rightarrow \threadstate s l & \quad \mbox{such that $\threadstate s l \in T$ ,} & \quad \mbox{and} \label{equation: additional edges 1} \\
  \threadstate s \ttdinitlocal \rightarrow \threadstate s l & \quad \mbox{such that $\threadstate s l \in T$ .}                    \label{equation: additional edges 2}
\end{eqnarray}

Suppose now some thread state $t_0 = \threadstate{s_0}{l_0}$ is reachable
in $\ttd[n]$, for some~$n$. Then there exists a path from some global state
$\state{\ttdshared[J]}{\range[]{l_1}{l_n}}$ such that $\threadstate
{\ttdshared[J]} {l_i} \in T$ for all $i$, to a global state with shared
component $s_0$ and some thread in local state~$l_0$. We can attach, to the
front of this path, the prefix
\begin{eqnarray*}
  \state{\ttdinitshared}{\range[]{\ttdinitlocal}{\ttdinitlocal}} & \wbox{$\transsymbol$} & \state{\underline{\ttdshared[J]}}{\underline{l_1},\ttdinitlocal,\range[]{\ttdinitlocal}{\ttdinitlocal}} \\
                                                        & \transsymbol & \state{           \ttdshared[J]} {l_1,\underline{l_2},\range[]{\ttdinitlocal}{\ttdinitlocal}} \\
                                                        &              & \quad \quad \cdots \\
                                                        & \transsymbol & \state{           \ttdshared[J]} {l_1,l_2,\range[]{l_3}{\underline{l_n}}} \ ,
\end{eqnarray*}
with the underlined symbols changed. The new path reaches $t_0$ in
$\ttd[n]'$.

Conversely, suppose some thread state $t_0 = \threadstate{s_0}{l_0}$ such
that $s_0 \not= \ttdinitshared$, $l_0 \not= \ttdinitlocal$ is reachable in
$\ttd[n]'$, for some~$n$. Then there exists a path $p'$ from
$\{\ttdinitshared\} \times \{\ttdinitlocal\}^n$ to a global state with shared
component $s_0$ and some thread in local state~$l_0$. The very first
transition of $p'$ is by some thread executing an edge of type
\equationref[]{additional edges 1}, since those are the only edges leaving
the unique initial state $\ttdinitstatepair$. Let that be thread number
$i$, and let $\threadstate s l \in T$ be the new state of thread $i$.

Consider now an arbitrary thread $j \in \{1,\ldots,n\} \setminus \{i\}$;
its local state after the first transition along $p'$ is $\ttdinitlocal$.
\begin{itemize}

\item If thread $j$ is never executed along $p'$, we build a new path $p''$
  by inserting edge $\edge s \ttdinitlocal s l$, executed by thread $j$, right
  after the first transition in~$p'$. This is a valid edge (of type
  \equationref[]{additional edges 2}) since $\threadstate s l \in T$. The
  edge moves thread $j$ into an initial thread state $(s,l) \in T$. The
  modified state sequence remains a valid path in $\ttd[n]'$ since no
  shared states have been changed, and thread $j$ is inactive henceforth.

\item If thread $j$ is executed along $p'$, then the first edge it executes
  must be of type \equationref[]{additional edges 2}, since again this is
  the only way to get out of local state $\ttdinitlocal$. Let
  $\threadstate{\bar s}{\bar l} \in T$ be the state of thread $j$ after
  executing this first edge. Then $\edge s \ttdinitlocal s {\bar l}$ is a
  valid edge (of type \equationref[]{additional edges 2}): from
  $\threadstate s l \in T$ and $\threadstate{\bar s}{\bar l} \in T$, we
  conclude $\threadstate s {\bar l} \in T$, by property~\equationref[]{box
    property}. We now build a new path $p''$, by removing from $p'$ thread
  $j$'s first transition, and instead inserting, right behind the first
  transition of $p'$, a transition where thread $j$ executes edge $\edge s
  \ttdinitlocal s {\bar l}$:
  \[
  \begin{array}{cccl}
    & p'   & :: & \edge [i] \ttdinitshared \ttdinitlocal s t \ , \quad \ldots \quad , \edge [j] {\bar s} \ttdinitlocal {\bar s} {\bar l} \\[1mm]
    \mbox{becomes} \\
    & p''  & :: & \edge [i] \ttdinitshared \ttdinitlocal s t \ , \edge [j] s \ttdinitlocal s {\bar l} \ , \quad \ldots
  \end{array}
  \]
  (here we add a thread index on top of an edge's arrow, to indicate the
  identity of the executing thread). The modified state sequence remains a
  valid path in $\ttd[n]'$, since the shared states ``match'' and are not
  changed by any of the removed or inserted edges. Moving the local state
  change of thread $j$ (from $\ttdinitlocal$ to $\bar l$) forward leaves
  the path intact, since the original edge $\edge {\bar s} \ttdinitlocal
  {\bar s} {\bar l}$ was thread $j$'s first activity.

\end{itemize}
This procedure is applied to every thread $j \not= i$, with the result
that, after the first $n$ transitions, all threads are in a state belonging
to $T$. The suffix of $p''$ following these transitions reaches $t_0$ in
$\ttd[n]$.\eop

\section{Proof of Lemma \lemmaref[]{counter values single iteration}}
\label{appendix: Proof of Lemma counter values single iteration}

\begin{REPEATLEM}{\thelemmaCounterValuesSingleIteration}
  Let $b_{\ttdlocal} = \Sigma_{\ttdlocal}(1)$ if $l_k = \ttdlocal$ (path
  $\expand \sigma$ ends in local state $\ttdlocal$), and $b_{\ttdlocal} =
  \Sigma_{\ttdlocal}(0)$ otherwise. Then $\Sigma_{\ttdlocal}(n_{\ttdlocal})
  = n_{\ttdlocal} \maxplus[b_{\ttdlocal}] \delta_{\ttdlocal}$ .
\end{REPEATLEM}
\Proof: by induction on the number $k$ of vertices of $\expand \sigma =
\range[]{t_1}{t_k}$.

\fbox{$k=1$:} then $\expand \sigma$ has no edges, so
$\Sigma_{\ttdlocal}(n_{\ttdlocal}) = n_{\ttdlocal}$, $b_{\ttdlocal} = 0$,
and $\delta_{\ttdlocal} = 0$. Thus, $\Sigma_{\ttdlocal}(n_{\ttdlocal}) =
n_{\ttdlocal} = n_{\ttdlocal} \maxplus[b_{\ttdlocal}] 0 = n_{\ttdlocal}
\maxplus[b_{\ttdlocal}] \delta_{\ttdlocal}$.

\fbox{$k \rightarrow k+1$:} Suppose $\expand \sigma =
\range[]{t_1}{t_{k+1}}$ has $k+1$ vertices, and \lemmaref{counter values
  single iteration} holds for all paths of $k$ vertices. One such path is
the \emphasize{suffix} $\expand \tau = \range[]{t_2}{t_{k+1}}$ of $\expand
\sigma$. By the induction hypothesis, $\expand \tau$'s summary function
$\Tau_{\ttdlocal}$ satisfies $\Tau_{\ttdlocal}(n_{\ttdlocal}) =
n_{\ttdlocal} \maxplus[c_{\ttdlocal}] \gamma_{\ttdlocal}$ for the real edge
summary $\gamma_{\ttdlocal}$ along $\expand \tau$, and $c_{\ttdlocal} =
\Tau_{\ttdlocal}(1)$ if $l_{k+1} = \ttdlocal$; otherwise $c_{\ttdlocal} =
\Tau_{\ttdlocal}(0)$. Note that $\expand \tau$ and $\expand \sigma$ have
the same final state $t_{k+1} = (s_{k+1},l_{k+1})$.

We now distinguish what \algorithmref{Symbolically executing a path for
  local state l} does to the first edge $e_1 = (t_1,t_2) =
((s_1,l_1),(s_2,l_2))$ of $\expand \sigma$ (which is traversed last):
\begin{description}

\item[Case 1:] $e_1 \in \ttdtranss$ and $l_1 = \ttdlocal$: Then
  $\Sigma_{\ttdlocal}(n_{\ttdlocal}) = \Tau_{\ttdlocal}(n_{\ttdlocal}) +
  1$, $\delta_{\ttdlocal} = \gamma_{\ttdlocal} + 1$, and $b_{\ttdlocal} =
  c_{\ttdlocal} + 1$. Using the induction hypothesis (IH), we get
  $\Sigma_{\ttdlocal}(n_{\ttdlocal}) = n_{\ttdlocal}
  \maxplus[c_{\ttdlocal}](\delta_{\ttdlocal} - 1) + 1$.
  \begin{itemize}

  \item If $n_{\ttdlocal} + \delta_{\ttdlocal} - 1 \atl c_{\ttdlocal}$,
    then $n_{\ttdlocal} \maxplus[c_{\ttdlocal}](\delta_{\ttdlocal} - 1) + 1
    = n_{\ttdlocal} + \delta_{\ttdlocal} = n_{\ttdlocal}
    \maxplus[b_{\ttdlocal}] \delta_{\ttdlocal}$ since $n_{\ttdlocal} +
    \delta_{\ttdlocal} \atl c_{\ttdlocal} + 1 = b_{\ttdlocal}$.

  \item If $n_{\ttdlocal} + \delta_{\ttdlocal} - 1 < c_{\ttdlocal}$, then
    $n_{\ttdlocal} \maxplus[c_{\ttdlocal}](\delta_{\ttdlocal} - 1) + 1 =
    c_{\ttdlocal} + 1 = b_{\ttdlocal} = n_{\ttdlocal}
    \maxplus[b_{\ttdlocal}] \delta_{\ttdlocal}$ since $n_{\ttdlocal} +
    \delta_{\ttdlocal} < c_{\ttdlocal} + 1 = b_{\ttdlocal}$.

  \end{itemize}

\item[Case 2:] $e_1 \in \ttdtranss$ and $l_2 = \ttdlocal$: This case is
  analogous to Case 1; for completeness, we spell it out. We have
  $\Sigma_{\ttdlocal}(n_{\ttdlocal}) = \Tau_{\ttdlocal}(n_{\ttdlocal}) -
  1$, $\delta_{\ttdlocal} = \gamma_{\ttdlocal} - 1$, and $b_{\ttdlocal} =
  c_{\ttdlocal} - 1$. Using the IH, we get
  $\Sigma_{\ttdlocal}(n_{\ttdlocal}) = n_{\ttdlocal}
  \maxplus[c_{\ttdlocal}](\delta_{\ttdlocal} + 1) - 1$.
  \begin{itemize}

  \item If $n_{\ttdlocal} + \delta_{\ttdlocal} + 1 \atl c_{\ttdlocal}$,
    then $n_{\ttdlocal} \maxplus[c_{\ttdlocal}](\delta_{\ttdlocal} + 1) - 1
    = n_{\ttdlocal} + \delta_{\ttdlocal} = n_{\ttdlocal}
    \maxplus[b_{\ttdlocal}] \delta_{\ttdlocal}$ since $n_{\ttdlocal} +
    \delta_{\ttdlocal} \atl c_{\ttdlocal} - 1 = b_{\ttdlocal}$.

  \item If $n_{\ttdlocal} + \delta_{\ttdlocal} + 1 < c_{\ttdlocal}$, then
    $n_{\ttdlocal} \maxplus[c_{\ttdlocal}](\delta_{\ttdlocal} + 1) - 1 =
    c_{\ttdlocal} - 1 = b_{\ttdlocal} = n_{\ttdlocal}
    \maxplus[b_{\ttdlocal}] \delta_{\ttdlocal}$ since $n_{\ttdlocal} +
    \delta_{\ttdlocal} < c_{\ttdlocal} - 1 = b_{\ttdlocal}$.

  \end{itemize}

\item[Case 3:] $e_1 \in \ttdtranssexpand \setminus \ttdtranss$ and $l_1 =
  \ttdlocal$: Then $\Sigma_{\ttdlocal}(n_{\ttdlocal}) =
  \Tau_{\ttdlocal}(n_{\ttdlocal}) \maxminus 1 + 1$, $\delta_{\ttdlocal} =
  \gamma_{\ttdlocal}$, and $b_{\ttdlocal} = c_{\ttdlocal} \maxminus 1 + 1$.
  Using the IH, we get $\Sigma_{\ttdlocal}(n_{\ttdlocal}) = n_{\ttdlocal}
  \maxplus[c_{\ttdlocal}] \delta_{\ttdlocal} \maxminus 1 + 1$.
  \begin{itemize}

  \item If $c_{\ttdlocal} \atl 1$, then $b_{\ttdlocal} = c_{\ttdlocal}$, so
    $n_{\ttdlocal} \maxplus[c_{\ttdlocal}] \delta_{\ttdlocal} \atl
    c_{\ttdlocal} \atl 1$, hence $n_{\ttdlocal} \maxplus[c_{\ttdlocal}]
    \delta_{\ttdlocal} \maxminus 1 + 1 = n_{\ttdlocal}
    \maxplus[c_{\ttdlocal}] \delta_{\ttdlocal} = n_{\ttdlocal}
    \maxplus[b_{\ttdlocal}] \delta_{\ttdlocal}$.

  \item If $c_{\ttdlocal} = 0$, then $b_{\ttdlocal} = 1$.
    \begin{itemize}

    \item If $n_{\ttdlocal} + \delta_{\ttdlocal} \atl 1$, then
      $n_{\ttdlocal} \maxplus[c_{\ttdlocal}] \delta_{\ttdlocal} \maxminus 1
      + 1 = n_{\ttdlocal} + \delta_{\ttdlocal} \maxminus 1 + 1 =
      n_{\ttdlocal} + \delta_{\ttdlocal} = n_{\ttdlocal}
      \maxplus[b_{\ttdlocal}] \delta_{\ttdlocal}$.

    \item If $n_{\ttdlocal} + \delta_{\ttdlocal} \atm 0$, then
      $n_{\ttdlocal} \maxplus[c_{\ttdlocal}] \delta_{\ttdlocal} \maxminus 1
      + 1 = c_{\ttdlocal} \maxminus 1 + 1 = 1 = n_{\ttdlocal}
      \maxplus[b_{\ttdlocal}] \delta_{\ttdlocal}$.

    \end{itemize}

  \end{itemize}

\item[Case 4:] none of the above. In this case $e_1$ has no impact on the
  path summary generated by \algorithmref{Symbolically executing a path for
    local state l}. Thus, $\Sigma_{\ttdlocal}(n_{\ttdlocal}) =
  \Tau_{\ttdlocal}(n_{\ttdlocal})$; in particular we have $b_{\ttdlocal} =
  c_{\ttdlocal}$ and $\delta_{\ttdlocal} = \gamma_{\ttdlocal}$. Further,
  $\Sigma_{\ttdlocal}(n_{\ttdlocal}) = \Tau_{\ttdlocal}(n_{\ttdlocal})
  \symbolcomment{IH} = n_{\ttdlocal} \maxplus[c_{\ttdlocal}]
  \gamma_{\ttdlocal} = n_{\ttdlocal} \maxplus[b_{\ttdlocal}]
  \delta_{\ttdlocal}$.\eop

\end{description}

\section{Proof of Theorem \theoremref[]{counter values}}
\label{appendix: Proof of Theorem counter values}

\begin{REPEATTHE}{\thetheoremCounterValues}
  Let superscript $\iterate{}$ denote $\iterator$ function applications.
  Then, for $\iterator \atl 1$,
  \begin{equation}
    \label{equation: counter values proof}
    \iterate{\Sigma_{\ttdlocal}}(n_{\ttdlocal}) = \Sigma_{\ttdlocal}(n_{\ttdlocal}) \maxplus[b_{\ttdlocal}] (\iterator-1) \cdot \delta_{\ttdlocal} \ .
  \end{equation}
\end{REPEATTHE}
\Proof: by induction on $\iterator$. For $\iterator=1$, the right-hand side
(rhs) of \equationref[]{counter values proof} equals
$\Sigma_{\ttdlocal}(n_{\ttdlocal}) \maxplus[b_{\ttdlocal}] 0 =
\Sigma_{\ttdlocal}(n_{\ttdlocal})$ since $\Sigma_{\ttdlocal}(n_{\ttdlocal})
+ 0 = \Sigma_{\ttdlocal}(n_{\ttdlocal}) \atl b_{\ttdlocal}$ by
\lemmaref{counter values single iteration}.

Now suppose \equationref[]{counter values proof} holds. For the inductive
step we obtain:
\begin{eqnarray}
  \iterate[(\iterator+1)]{\Sigma_{\ttdlocal}}(n_{\ttdlocal}) &                                                                   = & \Sigma_{\ttdlocal}(\iterate{\Sigma_{\ttdlocal}}(n_{\ttdlocal})) \nonumber \\
                                                    & \symbolcomment{IH}                                                = & \Sigma_{\ttdlocal}(\Sigma_{\ttdlocal}(n_{\ttdlocal}) \maxplus[b_{\ttdlocal}] (\iterator-1) \cdot \delta_{\ttdlocal}) \nonumber \\
                                                    & \symbolcomment{Lem.~\lemmaref[]{counter values single iteration}} = & (\Sigma_{\ttdlocal}(n_{\ttdlocal}) \maxplus[b_{\ttdlocal}] (\iterator-1) \cdot \delta_{\ttdlocal}) \maxplus[b_{\ttdlocal}] \delta_{\ttdlocal} \ . \label{equation: counter values proof aux}
\end{eqnarray}
We now distinguish three cases ($\explain[]{\ldots}$ below contains proof
step justifications):

(1) If $\delta_{\ttdlocal} \atl 0$:
\[
\begin{array}{cl}
      & \mbox{\equationref[]{counter values proof aux}} \\
  = \ & \explain{$(\iterator-1) \cdot \delta_{\ttdlocal} \atl 0$, $\Sigma_{\ttdlocal}(n_{\ttdlocal}) \atl b_{\ttdlocal}$, hence $\Sigma_{\ttdlocal}(n_{\ttdlocal}) + (\iterator-1) \cdot \delta_{\ttdlocal} \atl b_{\ttdlocal}$} \\
      & (\Sigma_{\ttdlocal}(n_{\ttdlocal}) + (\iterator-1) \cdot  \delta_{\ttdlocal}) \maxplus[b_{\ttdlocal}] \delta_{\ttdlocal} \\
  =   & \explain{$\delta_{\ttdlocal} \atl 0$} \\
      & (\Sigma_{\ttdlocal}(n_{\ttdlocal}) + (\iterator-1) \cdot \delta_{\ttdlocal}) + \delta_{\ttdlocal} \\
  = \\
      & \Sigma_{\ttdlocal}(n_{\ttdlocal}) + \iterator \cdot \delta_{\ttdlocal} \\
  =   & \explain{$\Sigma_{\ttdlocal}(n_{\ttdlocal}) + \iterator \cdot \delta_{\ttdlocal} \atl b_{\ttdlocal}$} \\
      & \Sigma_{\ttdlocal}(n_{\ttdlocal}) \maxplus[b_{\ttdlocal}] \iterator \cdot \delta_{\ttdlocal} \ ,
\end{array}
\]
the final expression being the rhs of \equationref[]{counter values proof},
for $\iterator$ replaced by $\iterator+1$.

(2) If $\delta_{\ttdlocal} < 0$ and $\Sigma_{\ttdlocal}(n_{\ttdlocal}) + (\iterator-1) \cdot \delta_{\ttdlocal} <
b_{\ttdlocal}$, then also $\Sigma_{\ttdlocal}(n_{\ttdlocal}) + \iterator \cdot \delta_{\ttdlocal} < b_{\ttdlocal}$, and:
\[
\begin{array}{cl}
      & \mbox{\equationref[]{counter values proof aux}} \\
  = \ & \explain{$\Sigma_{\ttdlocal}(n_{\ttdlocal}) + (\iterator-1) \cdot \delta_{\ttdlocal} < b_{\ttdlocal}$} \\
      & b_{\ttdlocal} \maxplus[b_{\ttdlocal}] \delta_{\ttdlocal} \\
  =   & \explain{$\delta_{\ttdlocal} < 0$} \\
      & b_{\ttdlocal} \\
  =   & \explain{$\Sigma_{\ttdlocal}(n_{\ttdlocal}) + \iterator \cdot \delta_{\ttdlocal} < b_{\ttdlocal}$} \\
      & \Sigma_{\ttdlocal}(n_{\ttdlocal}) \maxplus[b_{\ttdlocal}] \iterator \cdot \delta_{\ttdlocal} \ .
\end{array}
\]

(3) If finally $\delta_{\ttdlocal} < 0$ and
$\Sigma_{\ttdlocal}(n_{\ttdlocal}) + (\iterator-1) \cdot \delta_{\ttdlocal}
\atl b_{\ttdlocal}$, then \equationref[]{counter values proof aux} reduces
to $(\Sigma_{\ttdlocal}(n_{\ttdlocal}) + (\iterator-1) \cdot
\delta_{\ttdlocal}) \maxplus[b_{\ttdlocal}] \delta_{\ttdlocal}$. To get an
overview of what we need to prove, let
\[
\begin{array}{rclcrcl}
  X & = & \Sigma_{\ttdlocal}(n_{\ttdlocal}) + (\iterator-1) \cdot \delta_{\ttdlocal} \ , & \quad & X' & = & \Sigma_{\ttdlocal}(n_{\ttdlocal}) \ , \\
  Y & = & \delta_{\ttdlocal}                                   \ , &       & Y' & = & \iterator \cdot \delta_{\ttdlocal} \ .
\end{array}
\]
Then (the reduced) \equationref[]{counter values proof aux} equals $X
\maxplus[b_{\ttdlocal}] Y$, and the rhs of \equationref[]{counter values
  proof} equals $X' \maxplus[b_{\ttdlocal}] Y'$. Further, observe that $X +
Y = X' + Y'$. This implies that $X \maxplus[b_{\ttdlocal}] Y = X'
\maxplus[b_{\ttdlocal}] Y'$, which follows immediately by distinguishing
whether $X+Y \atl b_{\ttdlocal}$ or not. The equality $X
\maxplus[b_{\ttdlocal}] Y = X' \maxplus[b_{\ttdlocal}] Y'$ is what we
needed to prove.\eop

\section{Making Regular Expressions Alternation-Free}
\label{appendix: Making Regular Expressions Alternation-Free}

\begin{LEM}
  \label{lemma: Making Regular Expressions Alternation-Free}
  Let $\RETWO$ and $\RETHREE$ be regular expressions. Then
  $\kleene{(\RETWO \choice \RETHREE)} = \kleene{(\kleene \RETWO \concat
    \kleene \RETHREE)}$.
\end{LEM}
\Proof: We show a subset relationship in both directions.

\

1. $\mbox{LHS} \subset \mbox{RHS}$:
\begin{equation*}
  \begin{array}{ccccccl}
    \RETWO                              & \subset & \ \kleene\RETWO \   & \subset & \kleene\RETWO \concat \kleene\RETHREE             & \quad & \mbox{(properties of $\kleene{}$)} \\
    \RETHREE                            & \subset & \ \kleene\RETHREE \ & \subset & \kleene\RETWO \concat \kleene\RETHREE             &       & \mbox{(ditto)} \\
    \RETWO \choice \RETHREE             &         &                      & \subset & \kleene\RETWO \concat \kleene\RETHREE             &       & \mbox{(by the above two and set theory)} \\
    \kleene{(\RETWO \choice \RETHREE)} &         &                      & \subset & \kleene{(\kleene\RETWO \concat \kleene\RETHREE)} &       & \mbox{(monotonicity of $\kleene{}$)}
  \end{array}
\end{equation*}

\

2. $\mbox{RHS} \subset \mbox{LHS}$:
\begin{equation*}
  \begin{array}{ccccl}
    \RETWO                                              & \subset & \RETWO \choice \RETHREE                         & \quad & \mbox{(property of $\choice$)} \\
    \kleene\RETWO                                      & \subset & \kleene{(\RETWO \choice \RETHREE)}             & \quad & \mbox{(monotonicity of $\kleene{}$)} \\
    \kleene\RETHREE                                    & \subset & \kleene{(\RETWO \choice \RETHREE)}             &       & \mbox{(by symmetry)} \\
    \kleene\RETWO \concat \kleene\RETHREE             & \subset & \kleene{(\RETWO \choice \RETHREE)}             &       & \mbox{(property of $\kleene{}$: $x \in \kleene\RE \land y \in \kleene\RE \Rightarrow x \concat y \in \kleene\RE$)} \\
    \kleene{(\kleene\RETWO \concat \kleene\RETHREE)} & \subset & \kleene{(\kleene{(\RETWO \choice \RETHREE)})} &       & \mbox{(monotonicity of $\kleene{}$)} \\
    \kleene{(\kleene\RETWO \concat \kleene\RETHREE)} & \subset & \kleene{(\RETWO \choice \RETHREE)}             &       & \mbox{(idempotence of $\kleene{}$: $\kleene{(\kleene\RE)} = \kleene\RE$)}
  \end{array}
\end{equation*}
\eop

\section{Recurrences of Conjunctions as Conjunctions of Recurrences}
\label{appendix: Recurrences of Conjunctions as Conjunctions of Recurrences}

We show that replacing a recurrence of a conjunction by the conjunction of
the recurrences applied to the individual conjuncts overapproximates.
Formally:
\begin{LEM}
  Let $A$ and $B$ be binary relations. Then $\iterate{(A \intersection B)}
  \subset \iterate A \intersection \iterate B$.
\end{LEM}
\Proof: We first formalize our notion of ``recurrence''. Let $C$ be a
binary relation. The $\iterator$-fold recurrence $\iterate C$ is relation
$C$ composed with itself $\iterator$ times, i.e.\ the~set
\[
\begin{array}{cl}
  \iterate C \ = \ \mbox{\Large $\compos$}_{i=1}^{\iterator} C \ = \ \{(x,y) \st \exists \range[]{c_1}{c_{\iterator-1}}: & (x,c_1) \in C, \ (c_1,c_2) \in C, \ \ldots \ , \\
                                                                                                        & (c_{\iterator-2},c_{\iterator-1}) \in C, \ (c_{\iterator-1},y) \in C\} \ .
\end{array}
\]
From this definition it follows that the recurrence operator $\iterate{}$
is monotone: $C_1 \subset C_2 \limplies \iterate C_1 \subset \iterate C_2$.
Therefore:
\[
\begin{array}{ccccl}
  A \intersection B             & \subset & A                                   & \quad \quad & \mbox{(set theory)} \\
  \iterate{(A \intersection B)} & \subset & \iterate A                          &             & \mbox{(monotonicity of $\iterate{}$)} \\
  \iterate{(A \intersection B)} & \subset & \iterate B                          &             & \mbox{(symmetry)} \\
  \iterate{(A \intersection B)} & \subset & \iterate A \intersection \iterate B &             & \mbox{(set theory)}
\end{array}
\]
\eop

\section{Proof of Theorem \theoremref[]{soundness}}
\label{appendix: Proof of Theorem soundness}

\begin{REPEATTHE}{\thetheoremSoundness}
  If, for each quotient path $\quotient \sigma$ from $\ttdinitstate$'s to
  $\ttdfinalstate$'s SCC, \algorithmref{reachability} returns
  \emph{unreachable}, then $\ttdfinalstate$ is unreachable in $\ttdinf$.
\end{REPEATTHE}
\Proof: we show the contrapositive: if thread state $\ttdfinalstate$ is
reachable in $\ttdinf$, then there exists a path $\quotient \sigma$ in
$\ttdquotient$ from $\ttdinitstate$'s to $\ttdfinalstate$'s SCC such that,
for any regular expression encoding $\RE$ of $\quotient \sigma$,
$\LAND_{\ttdlocal \in \ttdlocals} \pathformula(\RE,l)$ is satisfiable. If
this is the case, \algorithmref{reachability} does not enter
\lineref{unreachable}. Since there is no other opportunity for the
algorithm to return \emph{unreachable along $\quotient \sigma$}, the
contrapositive is proved.

Suppose $\ttdfinalstate$ is reachable in $\ttdinf$, say via a path $p$ in
$\ttd[n]$ of the form
\begin{equation}
  \label{equation: witness path}
  p \ :: \ \state{\ttdinitshared}{\underbrace{\ttdinitlocal, \ldots, \ttdinitlocal}_n} \quad \transsymbol \quad \cdots \quad \transsymbol \quad \state \ttdfinalshared {l_1', \ldots, l_{y-1}', \ttdfinallocal, l_{y+1}', \ldots, l_n'} \ ,
\end{equation}
and let $(\range[]{e_1}{e_{|p|}}) \in \ttdtranss^{|p|}$ be the sequence of
TTD edges executed along~$p$. We first construct a path $\expand \sigma$
from $\ttdinitstate$ to $\ttdfinalstate$ in $\ttdexpand$,
by processing the $e_i$ as follows:
\begin{enumerate}[(1)]

\item Edge $e_1$ (which starts in $\ttdinitstate$) is processed by copying
  it to $\expand \sigma$.

\item Suppose edge $e_{i-1}$ has been processed, and suppose its target
  state is $(s,l)$. Edge $e_i$'s source state has shared component $s$ as
  well, since it is executed in $p$ from the global state reached after
  executing $e_{i-1}$. So let $e_i$'s source state be $(s,l')$.

  Edge $e_i$ is now processed as follows. If $l=l'$, append $e_i$ to
  $\expand \sigma$. Otherwise, first append $\expedge s l s {l'}$ to
  $\expand \sigma$, then $e_i$. Note that $\expedge s l s {l'}$ is a valid
  expansion edge in $\ttdtranssexpand$, since there exist two
  edges, $e_{i-1}$ and $e_i$, adjacent to the expansion edge's source and
  target, respectively.
\end{enumerate}
Step (2) is repeated until all edges have been processed. It is clear by
construction that $\expand \sigma$ is a valid path in $\ttdexpand$, and
that it starts in $\ttdinitstate = \ttdinitstatepair$. We finally have to
show that it ends in $\ttdfinalstate = \ttdfinalstatepair$. It may in fact
not: let $(\ttdfinalshared,l_f)$ be the target state of the final edge
$e_{|p|}$; $l_f$ may or may not be equal to $\ttdfinallocal$. If it is not,
we append an edge $\expedge \ttdfinalshared {l_f} \ttdfinalshared
\ttdfinallocal$ to $\expand \sigma$. This is a valid expansion edge by
\definitionref{expanded ttd}, and $\expand \sigma$ now ends in
$\ttdfinalstate$.

We observe of this construction that $\expand \sigma$ consists of all TTD
edges fired along $p$, in that order, plus possibly some expansion edges
inserted in between or at the end.
Let now $\quotient \sigma$ be the corresponding quotient path in
$\ttdquotient$ (it runs from $\ttdinitstate$'s to $\ttdfinalstate$'s SCC)
and $\RE$ a regular expression encoding of $\quotient \sigma$. We show
$\LAND_{\ttdlocal \in \ttdlocals} \pathformula(\RE,l)$ is satisfiable.

We begin by showing a relationship between formula $\pathformula(\RE,l)$
(over regular expressions with loops) and ``unwound'' expressions. We first
formalize the concept of expression unwinding. In contrast to~$\quotient
\sigma$, $\RE$ unambiguously identifies loops, via its Kleene star
subexpressions. Let therefore $\range[]{\Loop_1}{\Loop_m}$ be the
loops in $\RE$. Given non-negative integers
$\range[]{\iterator_1}{\iterator_m}$, the
\emph{$(\range[]{\iterator_1}{\iterator_m})$-unwinding} of $\RE$ is the
sequence of edges over $\ttdtranssexpand$ obtained by replacing each loop
$\Loop_i$, say of the form $\kleene{\re_i}$, by $\re_i \concat \ldots
\concat \re_i$, with $\iterator_i$ occurrences of~$\re_i$. By construction,
the $(\range[]{\iterator_1}{\iterator_m})$-unwinding of $\RE$ forms a path
in $\ttdexpand$.
\begin{LEM}
  \label{lemma: loop path summary and flat summary}
  Let $\range[]{\iterator_1}{\iterator_m} \in \NN$, and $\expand \tau$ be
  $\RE$'s $(\range[]{\iterator_1}{\iterator_m})$-unwinding. Let also
  $\ttdlocal$ be a local state, and $x = ( \ttdlocal = \ttdfinallocal \ ?
  \ 1 \ : \ 0)$. Let finally $\Tau_{\ttdlocal}$ be path $\expand
  \tau$'s summary function for local state $\ttdlocal$. Then the
  following formula is valid:
  \[
  \begin{array}{lcl}
    \Tau_{\ttdlocal}(x) \atl 1 \limplies \pathformula(\RE,l) & \quad & \mbox{if $\ttdlocal = \ttdinitlocal$, and} \\
    \Tau_{\ttdlocal}(x)  =   0 \limplies \pathformula(\RE,l) &       & \mbox{otherwise} \ .
  \end{array}
  \]
\end{LEM}
\Proof: $\pathformula(\RE,l)$ and the
summary function $\Tau_{\ttdlocal}$ are computed over the same path, except
that in the latter, each loop $\Loop_i$ has been unwound $\iterator_i$
times. By \theoremref{counter values}, the closed-form terms used in
$\pathformula(\RE,l)$ for innermost loops yield the same values as the
summaries of the unwound paths. Non-innermost loops are overapproximated by
$\pathformula(\RE,l)$, preserving the satisfaction of assignment given by
$\range[]{\iterator_1}{\iterator_m}$.\eop

By \lemmaref{loop path summary and flat summary}, in order to show that
$\LAND_{\ttdlocal \in \ttdlocals} \pathformula(\RE,l)$ is satisfiable, it
suffices to find $\range[]{\iterator_1}{\iterator_m} \in \NN$ such that,
for every $\ttdlocal \in \ttdlocals$, $\Tau_{\ttdlocal}(x) \atl 1$ if
$\ttdlocal = \ttdinitlocal$, and $\Tau_{\ttdlocal}(x) = 0$ otherwise, for
$\Tau_{\ttdlocal}$ as in the lemma. To this end, consider path $\expand
\sigma$ constructed above. Since expression $\RE$ captures all paths in
$\ttdexpand$ represented by quotient path $\quotient \sigma$, the
$\ttdtranssexpand$ sequence $\expand \sigma$ matches regular expression
$\RE$. Let therefore $\range[]{\iterator_1}{\iterator_m}$ be the numbers of
iterations of each Kleene star that witness the match. $\RE$'s
$(\range[]{\iterator_1}{\iterator_m})$-unwinding is exactly the summary
function $\Sigma_{\ttdlocal}$ of path~$\expand \sigma$. It remains to show
that $\Sigma_{\ttdinitlocal}(x) \atl 1$, and $\Sigma_{\ttdlocal}(x) = 0$
for $\ttdlocal \not= \ttdinitlocal$.

We have $\Sigma_{\ttdinitlocal}(x) \atl 1$ since backward-traversing the
first edge of path $p$ increments counter $n_{\ttdinitlocal}$ (the property
also holds in the trivial case that $p$ has no edges). The claim
$\Sigma_{\ttdlocal}(x) = 0$ for $\ttdlocal \not= \ttdinitlocal$ is more
involved; we prove it by generalization. Let $\expand \sigma$ be
arbitrarily decomposed into segments $\expand \rho \circ \expand \pi$, such
that $\expand \pi$ is any \emph{suffix} of $\expand \sigma$, with summary
function $\Pi_{\ttdlocal}$. Let global path $q$ be the suffix of $p$
``corresponding'' to $\expand \pi$, i.e.\ the suffix of $p$ starting after
all edges of $\expand \rho$ have fired. We show
\begin{equation}
  \label{equation: generalized claim}
  \Pi_{\ttdlocal}(x) \atm n_{\ttdlocal}(q_1), \mbox{\ for the initial state $q_1$ of $q$.}
\end{equation}
\Equationref{generalized claim} is sufficient for $\Sigma_{\ttdlocal}(x) =
0$: let $\expand \pi = \expand \sigma$, hence $q = p$. Then
\equationref[]{generalized claim} becomes $\Sigma_{\ttdlocal}(x) \atm
n_{\ttdlocal}(p_1)$ with $p_1 = \state{\ttdinitshared}{\ttdinitlocal,
  \ldots, \ttdinitlocal}$. Since $\ttdlocal \not= \ttdinitlocal$, we have
$n_{\ttdlocal}(p_1) = 0$, so $\Sigma_{\ttdlocal}(x) = 0$ follows.

We now prove $\Pi_{\ttdlocal}(x) \atm n_{\ttdlocal}(q_1)$ by induction on
the length of $\expand \pi$. If $\expand \pi$ is empty, then
$\Pi_{\ttdlocal}(x) = x$ (\algorithmref{Symbolically executing a path for
  local state l}), and $q$ is empty as well. Hence $q_1$ is the final state
of $p$. If $l = \ttdfinallocal$, then $x = 1$ and $n_{\ttdlocal}(q_1) \atl
1$, so $1 = x = \Pi_{\ttdlocal}(x) \atm n_{\ttdlocal}(q_1)$. If $\ttdlocal
\not= \ttdfinallocal$, then $x = 0$ and the property holds trivially.

Suppose now \equationref[]{generalized claim} holds for the suffix $\expand
\delta$ of $\expand \sigma$ equal to $\expand \pi$ except for the first
edge of $\expand \pi$. Call this edge $e$: $\expand \pi = \{e\} \circ
\expand \delta$.
\begin{itemize}

\item if $e$ is a real edge of $\expand \sigma$, then it is fired along
  $q$. Doing so increases counter $n_{\ttdlocal}$ if $e$ starts in local
  state $\ttdlocal$, it decreases $n_{\ttdlocal}$ if $e$ ends in local
  state $\ttdlocal$, and leaves $n_{\ttdlocal}$ invariant if not adjacent
  to $\ttdlocal$. These updates are in agreement with what the path summary
  function $\Pi_{\ttdlocal}$ does to its integer argument
  (\algorithmref{Symbolically executing a path for local state l}, first
  two \plkeyword{if} clauses). \Equationref{generalized claim} is thus
  preserved across $e$.

\item if $e = \expedge s j s {j'}$ is an expansion edge of $\expand
  \sigma$, then it of course does not exist in $q$ and thus does not affect
  $n_{\ttdlocal}$. If $l \not= j$, summary $\Sigma_{\ttdlocal}$ does not
  change either, by \algorithmref{Symbolically executing a path for local
    state l}, final \plkeyword{if} clause.

  If $l=j$, we note that $e$ cannot be the first edge of $\expand \sigma$:
  by construction, this first edge is a real edge. Since it is not the
  first, $e$ is preceded by a \emph{real} edge $e^- = \edge{\cdot}{\cdot} s
  j$ of $\expand \sigma$ that fired along $p$. This implies that the first
  state $q_1$ of $q$ contains a thread in local state $j$: $n_j(q_1) \atl
  1$.

  Let now $\Delta_j$ be $\expand \delta$'s summary function. Since $\expand
  \delta$ and $\expand \pi$ differ only by expansion edge $e$, by
  \algorithmref{Symbolically executing a path for local state l} tells us
  that $\Pi_j(x) = \Delta_j(x) \maxminus 1 + 1$, and by the induction
  hypothesis, $\Delta_j(x) \atm n_j(q_1)$. If now $\Delta_j(x) \atl 1$,
  then $\Pi_j(x) = \Delta_j(x)$, and $\Pi_j(x) \atm n_j(q_1)$ holds. If,
  however, $\Delta_j(x) = 0$, then also $\Pi_j(x) = 1 \atm n_j(q_1)$, which
  concludes the proof.\eop

\end{itemize}

\section{Correctness for the Simple-Loop Case}
\label{appendix: Correctness for the Simple-Loop Case}

In the following we show that, if all loops in $\ttdexpand$ are simple,
\algorithmref{reachability} is not only sound but also complete, i.e.\ it
never returns ``unknown''. The latter can happen in
\algorithmref{reachability} in two places: in \lineref{unknown 1} --- which
is inside the loop guarded by the condition ``$\RE$ contains loop nests''
and thus unreachable if all loops are simple --- and in \lineref{unknown
  2}. To show that \lineref{unknown 2} is also unreachable, we prove: if
the satisfiability check in \lineref{satisfiability check 1} is successful,
i.e.\ $\LAND_{\ttdlocal \in \ttdlocals} \pathformula(\RE,\ttdlocal)$ is
satisfiable with assignment $\range[]{\iterator_1}{\iterator_q}$, then
$\mbox{\textsc{Unwind}}(\RE, \range[]{\iterator_1}{\iterator_q})$
represents a feasible execution path.

\begin{THE}
  If there exists a path in $\ttdquotient$ from $\ttdinitstate$'s to
  $\ttdfinalstate$'s SCC with regular expression encoding $\RE$ such that
  $\LAND_{\ttdlocal \in \ttdlocals} \pathformula(\RE,\ttdlocal)$ is
  satisfiable, then thread state $\ttdfinalstate$ is reachable in
  $\ttdinf$.
\end{THE}
\Proof: Let $\quotient \sigma$ and $\RE$ be such a path in $\ttdquotient$
and regular expression, and let $\range[]{\iterator_1}{\iterator_q}$ be an
assignment satisfying $\LAND_{\ttdlocal \in \ttdlocals}
\pathformula(\RE,\ttdlocal)$. The procedure in \algorithmref{Constructing a
  global witness path from a path in ttdquotient} constructs a path $p$ in
$\ttdinf$ that ends in a state containing a thread in $\ttdfinalstate$.
\lineref{unwinding quotient sigma} first unwinds $\RE$ into $\expand \sigma
= \range[]{t_1}{t_k}$ in $\ttdexpand$; note that $t_1 = \ttdinitstate$,
$t_k = \ttdfinalstate$. Starting from global state
$\state{\ttdfinalshared}{\ttdfinallocal}$ (\lineref{initialize p}), the
procedure now traverses $\expand \sigma$ backwards. Intuitively, each real
edge is executed backwards. Each expansion edge is processed by adding, to
all states currently present in $p$, a thread in the \emph{source} local
state $l_i$ of the edge \emphasize{if} the current first state $p_1$ does
not already contain a thread in $l_i$, denoted $n_i(p_1)=0$ in \lineref{if
  we need to expand}.
\begin{algorithm}[htbp]
  \begin{algorithmic}[1]
    \REQUIRE path $\quotient \sigma$ in $\ttdquotient$, reg.\ expr.\ $\RE$, satisfying assignment $\range[]{\iterator_1}{\iterator_q}$
    \STMT let $\expand \sigma = \range[]{t_1}{t_k}$ be the $(\range[]{\iterator_1}{\iterator_q})$-unwinding of $\RE$ \label{line: unwinding quotient sigma} \COMMENT{$\threadstate{t_i}{t_{i+1}} \in \ttdexpand$}
    \STMT \code{$e_i$ := $(t_i,t_{i+1})$ for $1 \atm i < k$ , $\threadstate{s_i}{l_i}$ := $t_i$ for $1 \atm i \atm k$}
    \STATE \code{$p$ := "$\state{\ttdfinalshared}{\ttdfinallocal}$"} \label{line: initialize p}
    \FOR {\code{$i$ := $k-1$} \plkeyword{downto} 1}
      \STATE let $p_1$ be the current first state along $p$
      \IF[$e_i$ = real edge]{$e_i \in \ttdtranss$}
        \STATE let $p_0$ be the global state obtained by executing $e_i$ backwards from $p_1$ \label{line: exec backwards}
        \STATE add \code{"$p_0 \transsymbol$"} to the front of $p$
      \ELSIF[$e_i$ = expansion edge]{$n_i(p_1) = 0$} \label{line: if we need to expand}
        \STATE to \emph{every state} along the current $p$, add a thread in local state $l_i$ \label{line: expand global state}
      \ENDIF
    \ENDFOR
  \end{algorithmic}
  \caption{Constructing a global witness path $p$ in $\ttdinf$ from a path $\quotient \sigma$ in $\ttdquotient$}
  \label{algorithm: Constructing a global witness path from a path in ttdquotient}
\end{algorithm}

Formally, the algorithm maintains the following invariant:
\begin{PRO}
  \label{property: path reconstruction invariant}
  When edge $e_i = (\threadstate{s_i}{l_i},\threadstate{s_{i+1}}{l_{i+1}})$
  (real or expansion) is processed, the first global state $p_1$ along $p$
  satisfies $p_1 \covers \state{s_{i+1}}{l_{i+1}}$.
\end{PRO}
This property (proved below) ensures that the step in \lineref{exec
  backwards} is executable. As a result, $p$ is, at any time, a valid path
in $\ttdinf$: when processing a real edge, by executing it backwards,
\propertyref{path reconstruction invariant} guarantees that the added
global transition is valid. When processing an expansion edge, by adding a
thread in a fixed local state to all states currently present in $p$, we
preserve all global transitions in $p$, due to the monotonicity property of
$\covers$ and $\transsymbol$.

\Propertyref{path reconstruction invariant} follows from a simple inductive
argument. It holds for $i=k-1$, since $e_{k-1}$ ends in $t_k =
\ttdfinalstate = \ttdfinalstatepair$, which is initially the first state of
$p$ (\lineref{initialize p}). Consider processing edge $e_i$. If the
previous edge $e_{i+1}$ is a real edge, the property holds for $e_i$
because $e_{i+1}$ was executed backwards, resulting in a thread in state
$\threadstate{s_{i+1}}{l_{i+1}}$. If $e_{i+1}$ is an expansion edge and
$l_{i+1}$ was added to all states along $p$, then it was added to $p_1$,
and the property holds for $e_i$. If $l_{i+1}$ was not added, this is
because the then first state $p_1$ of $p$ already contains a thread in
local state $l_{i+1}$, and $p_1$ is unchanged. Since $s_{i+1} = s_{i+2}$
($e_{i+1}$ = expansion edge), the property holds for $e_i$, too.

By \lineref{initialize p}, it is clear that $p$ ends in a state covering
$\ttdfinalstate$: the last state can only be changed by adding threads in
certain local states (\lineref{expand global state}), which has no bearing
on the covering property. It remains to be shown that, when
\algorithmref{Constructing a global witness path from a path in
  ttdquotient} terminates, the first state $p_1$ of $p$ is initial,
i.e.\ of the form
$\state{\ttdinitshared}{\range[]{\ttdinitlocal}{\ttdinitlocal}}$.

State $p_1$'s shared component is $\ttdinitshared$ since $\expand \sigma$ begins in this
shared state. Thus, the last real edge processed sets the shared state to
$\ttdinitshared$ (if none, we have $\ttdinitshared = \ttdfinalshared$). As
for the local states, let $\ttdlocal \not= \ttdinitlocal$; we show
$n_{\ttdlocal}(p_1)=0$. Let $x$ be the number of threads in local state
$\ttdlocal$ in the \emph{last} state of $p$, i.e., $x = 1$ if $\ttdlocal =
\ttdfinallocal$, and $x = 0$ otherwise. By \lemmaref{loop path summary and flat
  summary}, $\Sigma_{\ttdlocal}(x) = 0 \limplies
\pathformula(\RE,\ttdlocal)$. Since the assignment
$\range[]{\iterator_1}{\iterator_q}$ satisfies
$\pathformula(\RE,\ttdlocal)$, we conclude $\Sigma_{\ttdlocal}(x) = 0$.

We finally show $\Sigma_{\ttdlocal}(x) = n_{\ttdlocal}(p_1)$, from which
$n_{\ttdlocal}(p_1) = 0$ follows as desired. We prove this by induction on
the number of edges of $\expand \sigma$. If $\expand \sigma$ has no edges,
then $\Sigma_{\ttdlocal}(x) = x$, which equals $n_{\ttdlocal}(p_1)$ by the
definition of $x$ and by $p_1 = \state{\ttdfinalshared}{\ttdfinallocal}$.
For the inductive step, we distinguish the different ways an edge $e_i$ is
processed in \algorithmref{Constructing a global witness path from a path
  in ttdquotient}:
\begin{itemize}

\item Processing a real edge $e_i$ of $\expand \sigma$ that starts in local
  state $\ttdlocal$ creates a new global state $p_0$ for $p$ with
  $n_{\ttdlocal}(p_0) = n_{\ttdlocal}(p_1) + 1$.
  This is in agreement with what the path summary function
  $\Sigma_{\ttdlocal}$ does to its integer argument
  (\algorithmref{Symbolically executing a path for local state l}).

\item Analogous reasoning applies to a real edge that ends in local state
  $\ttdlocal$.

\item A real edge not adjacent to local state $\ttdlocal$ leaves
  $n_{\ttdlocal}$ unchanged, as does $\Sigma_{\ttdlocal}$.

\item Processing an expansion edge $e_i$ that starts in local state
  $\ttdlocal$ changes the first state $p_1$ to $p_1'$ such that
  $n_{\ttdlocal}(p_1') = n_{\ttdlocal}(p_1) + 1$ if $n_{\ttdlocal}(p_1)=0$,
  and $n_{\ttdlocal}(p_1') = n_{\ttdlocal}(p_1)$ otherwise. That is exactly
  the semantics of the operation $\maxminus 1 + 1$ that the path summary
  function applies to its argument in this case (\algorithmref{Symbolically
    executing a path for local state l}).

\item Processing an expansion edge $e_i$ that does not start in local state
  $\ttdlocal$ does not affect counter $n_{\ttdlocal}$. The same is true for
  $\Sigma_{\ttdlocal}$, by \algorithmref{Symbolically executing a path for
  local state l}.\eop

\end{itemize}

\end{document}